\documentclass{article}

\usepackage[dblblindworkshop, final]{neurips_2026}
\workshoptitle{AIDaR}
\usepackage{graphicx}
\usepackage{tikz}
\usetikzlibrary{positioning,arrows.meta}
\usepackage[utf8]{inputenc}
\usepackage[T1]{fontenc}

\usepackage{url}
\usepackage{booktabs}
\usepackage{amsfonts}
\usepackage{nicefrac}
\usepackage{microtype}
\usepackage{xcolor}
\usepackage{amsmath}
\usepackage{placeins}
\usepackage{algorithm}
\usepackage{algpseudocode}
\usepackage{array}

\usepackage{amssymb}
\usepackage{hyperref}
\providecommand{\floatconts}[3]{#2\label{#1}#3}

\title{GraphEP: A Calibrated Data System and Benchmark for Learning Cardiac Electrophysiology}

\author{%
  Matteo Baldan \\
  University of Padova\\
  \texttt{matteo.baldan.7@studenti.unipd.it} \\
  \And
  Chiara D'Ercoli \\
  Sapienza University of Rome \\
  \texttt{cd956@cam.ac.uk} \\
  \AND
  James Rowbottom \\
  University of Cambridge \\
  \texttt{jr908@cam.ac.uk} \\
  \And
  Pietro Liò \\
  University of Cambridge \\
  \texttt{pl219@cam.ac.uk} \\
  \And
  Stefania Fresca \\
  University of Washington \\
  \texttt{sfresca@uw.edu} \\
}

\begin{document}
\maketitle

\begin{abstract}
High-fidelity simulations can produce \textit{complete} scientific data at spatiotemporal resolutions that are difficult to obtain experimentally, but simulator outputs are not automatically ready for scientific machine learning tasks. Cardiac electrophysiology (EP) makes this gap especially clear: samples live on anatomy-dependent, high-resolution volumetric meshes, propagation depends on anisotropic material fields, and clinically relevant events occupy different temporal scales. We introduce GraphEP, a calibrated data system that turns volumetric cardiac meshes and simulation protocols into condensed graph-native, full-field transmembrane-potential trajectories. GraphEP is released with 732 left-ventricular (LV) simulations from 20 source anatomies, spanning healthy and scarred tissue substrates at different severities and multiple pacing configurations. Each sample packages geometry, material and tissue features, stimulation, graph connectivity, derived activation targets, and provenance in a PyTorch Geometric representation. We use GraphEP to compare four styles of inductive bias for full-field transmembrane-potential prediction $V_m(t)$: local message passing, global neural operators, multi-scale hierarchical models (hierarchical graph transformers (HGT)), and physiology-informed models. The physiology-informed family contains two complementary formulations: HGT-Lag factorises trajectories into activation timing and action-potential morphology, while the stateful Graph Aliev--Panfilov Action-Potential ODE (GAP$^2$ODE) evolves excitation and recovery autoregressively under frame-wise stimulation and a graph diffusion operator. We show wider spatial communication outperforms local message passing but incorporating physiological structure is the strongest inductive bias, with the two physiology-informed formulations leading on complementary metrics through explicit representation of propagation, recovery, and temporal state. GraphEP contributes both a reproducible route from calibrated mechanistic simulation in Cardiac EP to analysis-ready scientific data and a controlled setting for studying how local, global, multi-scale, and physiological inductive biases interact on irregular anatomical geometries.

\end{abstract}

\section{Introduction}

Cardiac electrophysiology (EP) couples fast, nonlinear cellular dynamics with anisotropic propagation through a three-dimensional myocardium. A simulation is therefore neither a conventional image nor a fixed-sensor time series: every anatomy is represented by a different unstructured volumetric mesh, and every vertex carries a signal characterised by a sharp depolarisation front followed by substantially slower recovery. Moreover, the domain is geometric rather than merely discrete, as conduction depends on local material properties such as fibre, sheet, and sheet-normal directions. Cardiac EP, therefore, offers a useful test bed to learn multi-scale spatiotemporal dynamics on irregular geometries.

The challenge for scientific machine learning (SciML) is not only to solve or approximate the underlying cardiac forward model, but also to transform simulation outputs into a reproducible and analysis-ready data representation. High-fidelity EP cohorts remain expensive to generate and difficult to compare. Therefore, neural surrogates commonly use a fixed or simplified geometry, and simulation outputs are rarely released with reusable graph features \cite{gsell2024forcepss}. A cross-anatomy dataset requires consistent geometry processing, physiological calibration, tissue labelling, stimulation protocols, graph construction, feature semantics, target extraction, normalisation, dataset splits, and provenance. Existing components such as openCARP~\cite{opencarp}, the ForCEPSS calibration protocol~\cite{gsell2024forcepss}, and public four-chamber human heart geometries~\cite{strocchi4chamber} make high-fidelity simulation possible, but are not, on their own, sufficient to define a reusable SciML corpus or benchmark.
The quantitative benchmark in this first release is restricted to LV samples. Its split is sample-disjoint but not anatomy-disjoint: all 20 anatomies occur across train, validation, and test, while pacing configurations and pathological manifestations vary between samples. The benchmark, therefore, evaluates the learning of propagation rules across unseen samples on known anatomical identities. In this contribution, we also show biventricular and whole-heart simulations to demonstrate the scope of the generation pipeline as a system ready for SciML data preparation.

We introduce GraphEP as three connected contributions: \emph{(i)} a reproducible path from three-dimensional meshes of cardiac geometries to calibrated solutions of the Monodomain partial differential equation (PDE) \cite{niederer2011benchmark} and graph-native samples; \emph{(ii)} an audited collection of 732 LV PyG geometric tensors spanning healthy and scarred substrates, Purkinje configurations, pacing sites, and non-resting boundary states; and \emph{(iii)} a controlled benchmark of local, global, multi-scale, and biophysically-inspired architectures for full-field voltage regression and autoregressive state evolution.

\begin{figure}[htpb]
\centering
  \caption{\small{\textbf{GraphEP samples.} Healthy and medium-scar examples over three paced beats. Heatmaps order nodes by first-beat activation; dashed lines mark successive stimuli. Scar delays and fragments propagation, leave late regions depolarised as the next stimulus arrives. Spatial snapshots show the corresponding wavefront moving around the scar obstacle.}}%
  \label{fig:corpus-example}%
  {\begin{minipage}[t]{0.42\linewidth}
    \centering\textbf{Healthy}\\
    \includegraphics[width=0.9\linewidth]{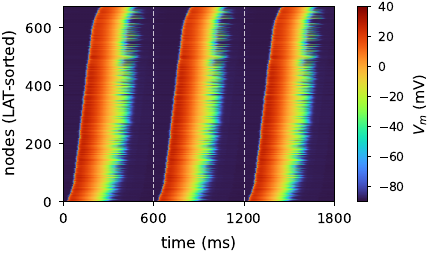}
    \includegraphics[width=0.9\linewidth]{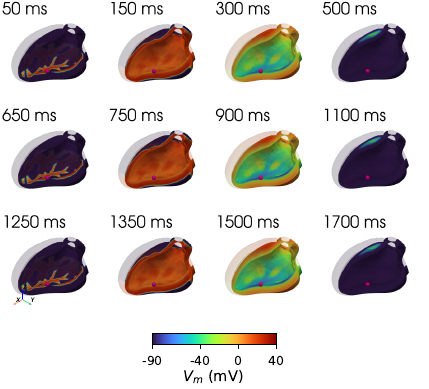}
  \end{minipage}
  \begin{minipage}[t]{0.42\linewidth}
    \centering\textbf{Medium scar}\\
    \includegraphics[width=0.9\linewidth]{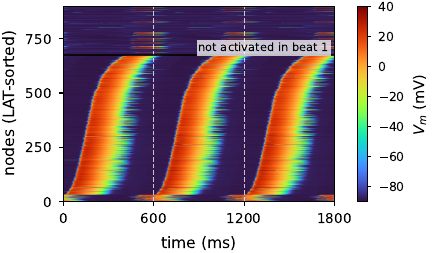}
    \includegraphics[width=0.9\linewidth]{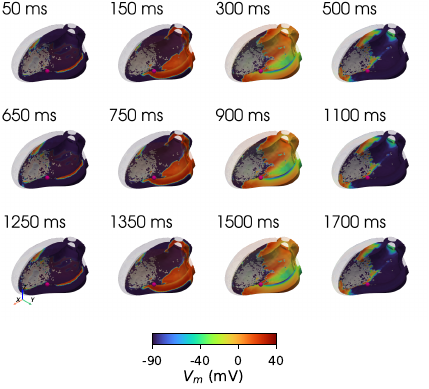}
  \end{minipage}}
\end{figure}

% \begin{wrapfigure}{r}{0.5\textwidth}
%   \vspace{-14pt}
%   \centering
%   \includegraphics[width=0.48\textwidth]{figures/cameraready/gap2ode_fig1/idx42_wrap/pick_42_cutaway_gt_pred_error.pdf}
%   \caption{\small GAP$^2$ODE on a held-out large-scar sample (71\% of tissue
%   scar-affected): ground truth, prediction, and absolute error on the native
%   mesh at times spanning the first beat's activation. Sample MAE
%   $8.3\,\mathrm{mV}$ against a $10.2\,\mathrm{mV}$ test mean; error
%   concentrates at the moving front while the wave negotiates the lesion.}
%   \label{fig:gap2ode-showcase}
%   \vspace{-10pt}
% \end{wrapfigure}

\section{Data and Methods}
\label{sec:data-methods}

GraphEP transforms anatomically realistic cardiac meshes into graph-structured EP time series through a sequence of auditable stages: geometry preparation and annotation, physiological calibration, graph construction, and optional coarsening. The pipeline was conceived with the aim of retaining the information needed to trace a machine learning sample back to its underlying geometry, material configuration, stimulation protocol, and state of the system at any point in time.

\subsection{Data generation and physiological calibration}
The pipeline has four stages: extraction and augmentation of cardiac geometries, physiological calibration, solving a PDE system with a finite element method (FEM) solver, and conversion into PyTorch Geometric representations. It accepts volumetric meshes as input, and it extracts selected cardiac regions; in the released dataset, we extract the LV from the public four-chamber cohort of~\cite{strocchi4chamber}. In this case, cardiac meshes are re-meshed to the target resolution and annotated with Universal Ventricular Coordinates ~\cite{uvc}. A rule-based construction assigns fibre, sheet, and sheet-normal directions, producing the local material frame used for anisotropic conduction. Endocardial fast-conduction trees provide rapid activation, while biased random walks over the element graph generate minor, medium, and large scar patterns. Each scar core is inactive and nearly insulating, whereas its excitable border zone is assigned a reduced directional conductivity. Details of the algorithms and mesh handling steps are reported in the supplementary material in Section \ref{sec:supp-mesh-handling}.

The ToR-ORd ventricular cell model~\cite{tomektorord} is first brought to a periodic steady state. Tissue conductivities are then calibrated following the ForCEPSS workflow~\cite{gsell2024forcepss}, which adapts the diffusion parameters to reproduce the prescribed conduction velocities at the chosen discretisation. 
Calibration is part of the data-generation pipeline itself rather than an external preprocessing step: the cellular model is initialised from periodic steady state and conductivities are tuned before cohort simulation.
A compact quantitative QC table is reported in supplementary material in the extended Table \ref{tab:graphep-lv-qc}. 

OpenCARP~\cite{opencarp} solves the Monodomain reaction--diffusion equation~\cite{monodomaincolli}. The solver is configurable following ForCEPPS recommendations and allows defining a stimulation protocol. In the released dataset we define three stimuli, each spaced $600\,\mathrm{ms}$ apart. We retain node-wise transmembrane voltage $V_m$ and derive first-threshold-crossing activation times. The frozen release stores 361 frames at $5\,\mathrm{ms}$ spacing over $1.8\,\mathrm{s}$; finer solver outputs, calibration parameters, and pathology-generation details are documented in the supplementary material. The results are then saved in ZARR format, allowing for efficient indexing, reading, and writing times to storage. This also enables simulation compression, considering that the majority of the dense node-voltage time-series retain resting potential values, thus allowing us to save memory.

\subsection{Graph-structured time-series representation}
GraphEP converts each finite element simulation into a graph representation suitable for geometric learning. To reduce training costs, the finite element graph is coarsened with a topology-aware first-order $k$-maximal-independent-set selector (\texttt{kmis}): native vertices are adjacent when they occur in a common tetrahedron. Each simulation becomes a PyTorch Geometric \texttt{Data} object with core tensors
\begin{equation}
\begin{aligned}
\mathtt{x}&\in\mathbb R^{n_i\times17}, &
\mathtt{pos}&\in\mathbb R^{n_i\times3}, &
\mathtt{edge\_index}&\in\mathbb N^{2\times M_i},\\
\mathtt{edge\_attr}&\in\mathbb R^{M_i\times5}, &
\mathtt{y},\mathtt{stimulus}&\in\mathbb R^{n_i\times T}, &
\mathtt{times}&\in\mathbb R^T,\\
\mathtt{y\_lat}&\in\mathbb R^{n_i\times1}, &
\mathtt{lat\_valid}&\in\{0,1\}^{n_i}, &
\mathtt{node\_volume}&\in\mathbb R_+^{n_i}.
\end{aligned}
\label{eq:dataspace}
\end{equation}
Here $n_i$ and $M_i$ vary with anatomy, and $T=361$ where $i=1,...,N$ and $N$ are the number of samples. Voltage and stimulus are stored in half precision (FP16), whereas geometry, features, distances, and activation targets use single precision (FP32).

The node channels are $[\mathbf f,\mathbf s,\boldsymbol\sigma,b^{\rm stim},t^{\rm stim},c,\mathbf q^{\rm UVC},d^{\rm aniso}]$, representing fibre and sheet directions, three intracellular conductivities, the stimulus mask and onset, the canonical tissue class, four UVC coordinates, and the anisotropic distance from the pacing region, respectively. Continuous channels are cluster-averaged and normalised; the stimulus mask uses logical OR, while the tissue class is inherited from the representative node. 

For an edge $(u,v)$, the five channels comprise: displacement projected onto the local fibre, sheet, and normal unit directions, Euclidean length, and conductivity projected along the edge. Hence, the representation retains the local material frame rather than reducing geometry to Cartesian distance alone.

The target \texttt{y} contains node-wise $V_m$ over three beats. The first valid rising crossing of $-40\,\mathrm{mV}$ defines \texttt{y\_lat}, and invalid nodes are masked by \texttt{lat\_valid}. Auxiliary fields include Euclidean, graph-geodesic, and conductivity-weighted stimulus distances, edge travel cost, tissue tags, scar points, protocol parameters, and provenance metadata. Each tetrahedron contributes one quarter of its volume to each of its four vertices, so \texttt{node\_volume} approximates the local integration measure associated with each graph node.

Individual samples from the corpus differ in anatomy, discretisation and activation patterns, so no node-wise correspondence exists across the corpus; it can instead be partitioned by indicators derived from the released data: scar severity, Purkinje presence, prepacing state at $t{=}0$, and activation extent (first-beat LAT fraction and peak active fraction), as demonstrated in the dataset summary of Fig.~\ref{fig:dataset-summary}.

% Additional \textbf{biophysical features} include the initial-voltage field, an initial-active mask, Purkinje-presence indicators and beat-local activation masks. Operator enrichment adds signed distance to the scar interface and core/border-zone region, together with the sample-specific Galerkin-coarsened finite element operator
% \begin{equation}
%     K_c=P^{\top} K P,
%     \label{eq:coarse-fem-main}
% \end{equation}
% %where $K$ is assembled from the native mesh and calibrated conductivity tensor, and $P$ maps native vertices to baked graph nodes. Edge coefficients and consistently aggregated nodal masses are attached as FEM stiffness to each edge and FEM mass to each node. These are offline, provenance-carrying sidecars; time since depolarisation, diastolic interval and voltage history are instead online state maintained by autoregressive models.
% where $K \in \mathbb{R}^{n_v \times n_v}$ is assembled on the native mesh from the calibrated conductivity tensor, and $P \in \{0,1\}^{n_v \times n_i}$ assigns native vertices to graph nodes. The lumped mass is coarsened by the same assignment, $M_c = P^\top M P$; graph edges carry the entries of $K_c$, nodes the diagonal of $M_c$. These graph features are precomputed offline; time since depolarisation, diastolic interval (DI) and voltage history are instead online state maintained by autoregressive models.

We include additional \textbf{biophysical features} that expose each sample's state and physics. Distance features: signed graph-geodesic distance to the scar interface and to the core/border-zone region, and conductivity-weighted distances to the lesion and to the Purkinje network (in addition to the Purkinje tissue class type flag). Operator features: the sample-specific Galerkin-coarsened finite element operator
\begin{equation}
    K_c = P^{\top} K P,
    \label{eq:coarse-fem-main}
\end{equation}
where $K \in \mathbb{R}^{n_v \times n_v}$ is assembled on the native mesh from the calibrated conductivity tensor, and $P \in \{0,1\}^{n_v \times n_i}$ assigns native vertices to graph nodes. The lumped mass is coarsened by the same assignment, $M_c = P^{\top} M P$; graph edges carry the entries of $K_c$, nodes the diagonal of $M_c$ (the physical diffusion operator and integration measure of the coarse graph). These features are precomputed offline; the time since depolarisation, diastolic interval (DI), and voltage history are instead online states maintained by autoregressive models.

\subsection{Prediction task and evaluation protocol}
We formulate the supervised learning problem as the approximation of the solution operator that maps anatomy, material properties, and pacing conditions to the resulting spatiotemporal transmembrane potential field. Conceptually, for a cardiac domain $\Omega_i$:
\begin{equation}
    \mathcal{F}:(\Omega_i,\, X_i(\mathbf{x}),\,S_i(\mathbf{x},t)) \longmapsto {V}_{m,i}(\mathbf x,t),
\end{equation}

where $\mathbf{x}\in\Omega_i$ denotes the spatial position, $X_i(\mathbf{x})$ represents the spatially varying geometric and material properties  and $S_i(\mathbf{x},t)$ describes the stimulation protocol for each sample $i=1,...,N$. In practice, these fields are discretised with the graph $\mathcal{G}_i$. We let $X_i$ be the static node and edge attributes described in the previous section, including geometry, local material orientation, tissue properties, and conductivity. This results in the discrete regression problem: 

\begin{equation}
\mathcal{F}_\theta: (\mathcal{G}_i,X_i,S_i) \longmapsto \widehat{\mathbf V}{m,i}, \qquad \widehat{\mathbf V}_{m,i}\in\mathbb{R}^{n_i\times T},
\end{equation}

where $n_i$ is the number of graph nodes, $T=361$ is the number of simulated voltage frames, and $\theta$ denotes the learnable model parameters.
The same regression problem can also be framed as an autoregressive task, where the operator does not learn the mapping from the initial conditions and static properties to the spatiotemporal transmembrane field, but instead maps the state of the system, conditioned by its circumstances, to a later time frame.

In addition to these direct solution-operator models, we evaluate a stateful graph neural ODE that instead learns the temporal evolution of a latent cardiac state.

The same regression problem can alternatively be framed as an autoregressive task. Rather than mapping the static inputs directly to the complete voltage trajectory, the model learns a state-transition operator that maps the current state of the system to a future state, conditioned on the graph, material properties, and stimulation:

\begin{equation}
\mathcal{F}_\theta: (\mathcal{G}_i,X_i,S_i, z_i(t)) \longmapsto z_i(t+\Delta t), \qquad \widehat{\mathbf V}_{m,i}(t)=\mathcal{R}(z_i(t)),
\end{equation}

where $z_i(t)$ denotes the state of the system at time $t$, which may include the transmembrane potential together with additional latent variables, and $\mathcal{R}_{\theta}$ decodes this state to the predicted voltage. The complete trajectory is obtained by recursively applying the transition operator over the $T$ time points.

The frozen split is stratified over samples by anatomy and scar severity, although it is not anatomy-disjoint. All 20 heart models, therefore, occur across the split, while held-out samples differ in pacing configuration and pathological development. The release contains 732 PyG samples, partitioned into 513 training, 146 validation, and 73 test samples, and stored in a frozen manifest with SHA-256 verification. Samples are packaged deterministically, and the archive-level verification procedure and storage budget are reported in the supplementary material. 

Each model is trained for 4,000 optimiser updates with pointwise voltage MSE and a temporal-variance term that discourages collapse to a flat trajectory. The checkpoint with the lowest validation node--time MAE is retained. Evaluation combines generic regression error with physiological endpoints: active-myocardium MAE. depolarisation amplitude MAE and ratio, local activation-time (LAT) MAE, action potential duration at 90\% repolarisation (APD90) MAE, and repolarisation time (RT) MAE. This distinction is deliberate: pointwise voltage error alone does not establish that a surrogate can reproduce physiologically meaningful timing or waveform morphology. Additional details are reported in the supplementary material.

\subsection{Model families and receptive-field axis}
We organise the benchmark as a progression from generic spatial approximation to increasingly structured representations of cardiac propagation. GATv2~\cite{brody2021attentive} provides a local message-passing baseline, while GINO~\cite{gino} and GWNO introduce domain-wide communication through global neural operators. GHR~\cite{carotti2026graph} and HGT then combine information across multiple spatial resolutions, preserving local anatomical detail while communicating over the whole ventricle. The proposed physiology-informed models add structure specific to electrophysiology: HGT-Lag separates propagation timing from action potential morphology, whereas the stateful graph ODE evolves excitation and recovery through a numerical integrator coupled to the sample-specific diffusion operator. This progression allows us to examine not only how far information travels through a model, but also how explicitly the model represents the mechanisms that generate the trajectory.

\subsection{Physiology-informed and multiscale models}
\vspace{-6pt}
\paragraph{Hierarchical Graph Transformer (HGT) and HGT-Lag.}
HGT follows a U-Net-like multi-resolution design and learns a map between continuous functions, thus describing a neural graph transformer operator. Node features are projected into a common latent space and progressively pooled onto regular 3-D grids of decreasing resolution. At each resolution, the spatially aggregated features are processed by linear-attention transformer blocks, which provide global information exchange at that spatial scale at a linear cost (number of tokens). The decoder maps coarse representations back to finer resolutions through inverse-distance-weighted interpolation from the pooled centroids, concatenates the corresponding encoder representations through skip connections, and refines the result with additional transformer blocks. The model, therefore, combines fine geometric detail with communication over progressively larger spatial scales. In the current implementation, graph edges are not used inside the transformer hierarchy. Instead, spatial neighbourhoods are induced by geometric pooling and interpolation.

The two variants share the same encoder and differ only in how the latent field is turned into a voltage trajectory. HGT applies a pointwise output network that regresses all frames of the trajectory directly. HGT-Lag, instead, uses a parametric action-potential decoder: from each node embedding, it predicts an amplitude, a baseline voltage, a local conduction speed, and a bounded residual correction to the activation time. A node's activation time is the geodesic distance to the stimulation site divided by its predicted conduction speed, plus a residual and a shared action-potential template that describes an upstroke of learnable width multiplied by a learnable plateau-to-repolarisation tail. This template is shifted by learnable activation time and scaled by the amplitude to reconstruct the entire trajectory. Because the template is a function of physical time from activation, the same decoder is evaluable on any temporal grid. This formulation introduces a physiological inductive bias by approximately factorising the prediction problem into \emph{when the wave arrives} and \emph{what local waveform follows}, and it exposes an explicit per-node activation time that can be directly supervised.

Layer updates, grid pooling and unpooling, the linear-attention block, and the two decoders are defined in Supplementary Section~\ref{sec:supp-hgt}, together with an outline of the model's architecture in Figure  \ref{fig:hgt_arch}.

\vspace{-6pt}
\paragraph{Physiology-informed Graph ODE}
% Unlike the previous models, the Physiology Informed graph ODE advances an excitable state of each node of the graph and recursively construct the voltage trajectories. The nodes are described by state variables that interpret the physiological behaviour of the cardiac muscle cell using the Aliev-Panfilov reaction. Each node carries normalised excitation $u_i$, recovery $w_i$ and latent memory. Propagation is governed by the sample-specific coarsened finite-element diffusion operator, while a lightweight GNN predicts bounded corrections to reaction rates and edge conductance. Stimulation is read at every frame and is gated by recovery state, so pacing acts as a time-dependent intervention on the evolving system. An implicit diffusion step stabilises integration on irregular meshes, and a learnt observation map reconstructs detailed voltage morphology without feeding its correction back into the physical state. The model therefore combines a prescribed excitable-dynamics scaffold with learnt, geometry-conditioned closure terms. Its equations, parameter bounds, online state, observation map and integration scheme are given in Supplementary Section~\ref{sec:supp-graph-ode}.

Unlike the previous models, the physiology-informed Graph Aliev--Panfilov Action-Potential ODE (GAP$^2$ODE) advances an excitable state on the graph using a neural reaction--diffusion ODE and constructs the voltage trajectory recursively. Each node carries normalised excitation $u_i$, recovery $w_i$ and a small latent memory, evolving under Aliev--Panfilov reaction dynamics; propagation is governed by the sample-specific coarsened finite-element operator of Eq.~\eqref{eq:coarse-fem-main}. A graph network encodes local geometry and tissue class once per sample, and its output sets per-node reaction rates and per-edge conductance gains within fixed physiological bounds, acting as a learned closure model for what coarsening and the phenomenological cell model omit. The stimulus enters the excitation equation at every frame through a recovery-dependent gate, so refractory tissue cannot be re-excited and pacing acts on the evolving state. An implicit diffusion step keeps integration stable on irregular meshes. Finally, an observation map decodes the state to transmembrane voltage: an affine map plus a bounded learned correction capturing waveform detail beyond the two-variable dynamics. Equations, parameter bounds, online state and integration scheme are given in Supplementary Section~\ref{sec:supp-graph-ode}.

\section{Results}

\subsection{Dataset scale, diversity and physiological structure}
\begin{figure}[htbp]
    \centering
    \includegraphics[width=0.81\textwidth]{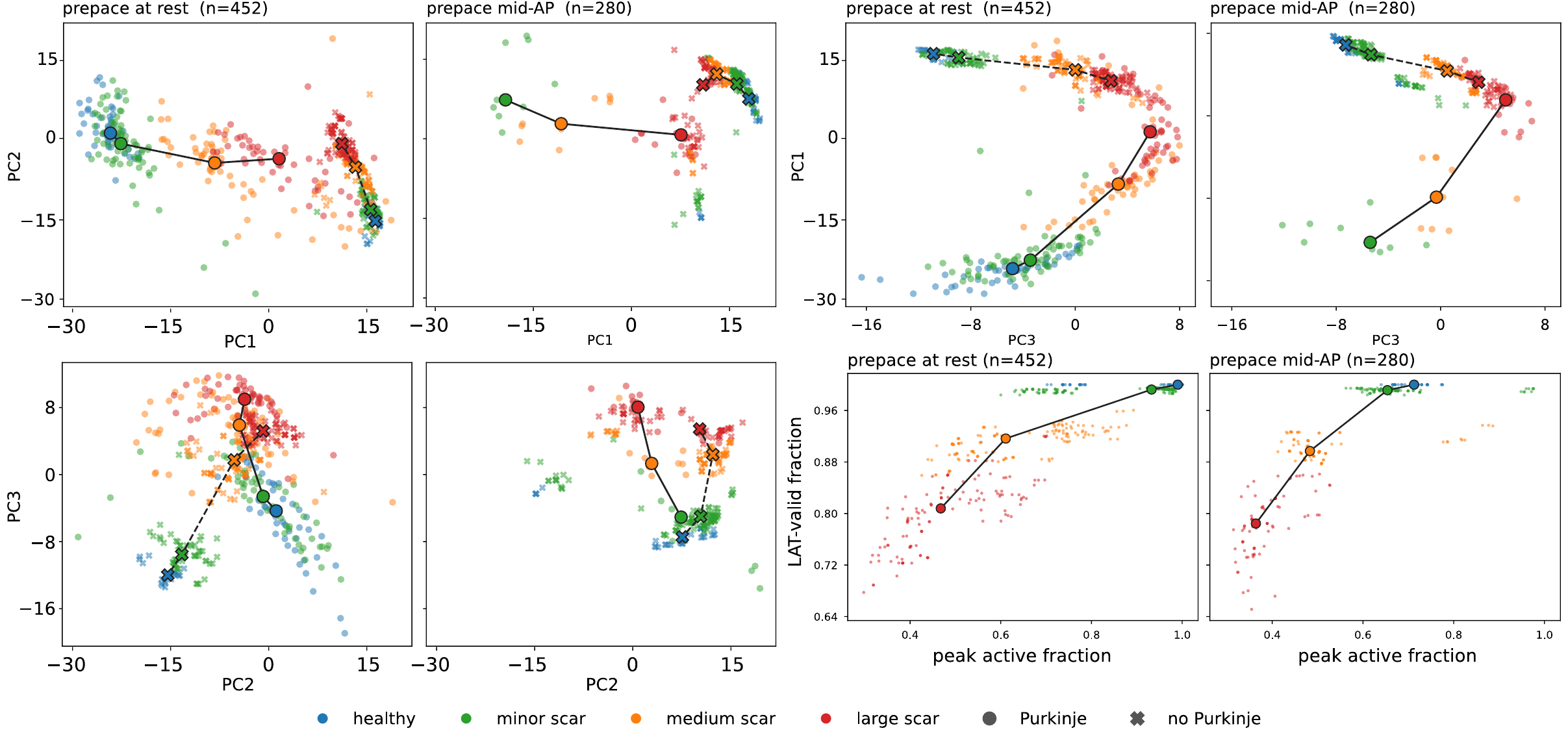}
    \caption{\small{Corpus variability. Dots are coloured by scar severity and marked by presence of the Purkinje network. Panels show pairwise projections of the first three principal components of model-facing $V_m$ signatures and the relation between first-beat LAT fraction and peak active fraction. Large markers denote group centroids. The projections illustrate that known controlled factors in the generation process are reflected in the resulting dynamics.}}
    \label{fig:dataset-summary}
\end{figure}

The released dataset contains 732 simulations from 20 source anatomies, each with 361 voltage frames at $5\,\mathrm{ms}$, spaced over three paced beats ($0$--$1800\,\mathrm{ms}$; BCL $600\,\mathrm{ms}$). The cohort includes healthy myocardium with minor, medium, and large random-walker scar patterns and multiple stimulus locations. Figure~\ref{fig:dataset-summary} summarises variability in model-facing $V_m$ signatures. The first three principal-component projections are associated with controlled cohort factors, including Purkinje configuration, boundary state, pacing location, and scar severity. We use these projections as a descriptive view of corpus variability rather than assigning a single causal factor to each component.

\subsection{Representative full-field dynamics}
Figure~\ref{fig:corpus-example} contrasts healthy and scarred samples from the same LV anatomy and pacing protocol. In the healthy example, the wavefront crosses the ventricle regularly, and the myocardium recovers before the following stimulus. In the matched scarred example, propagation around the lesion is slower and spatially heterogeneous, leaving some regions depolarised when the subsequent stimulus arrives at the $600\,\mathrm{ms}$ pacing boundary. This interaction between propagation, waveform morphology, and successive pacing cycles illustrates why full-field $V_m(t)$ might overlap excitable waves and create large delays.
\vspace{-6pt}
\paragraph{Extension beyond the LV benchmark.}
The same generation system supports biventricular tissue with independently generated LV and Right Ventricle (RV) Purkinje regions and whole-heart simulations coupling Courtemanche atrial dynamics~\cite{courtemanche} with ToR-ORd ventricles. These experiments demonstrate pipeline extensibility to a large variety of experiments on cardiac meshes.

\subsection{Benchmark performance and model inductive bias}
\begin{figure}[t]
  \centering
  \includegraphics[width=\textwidth]{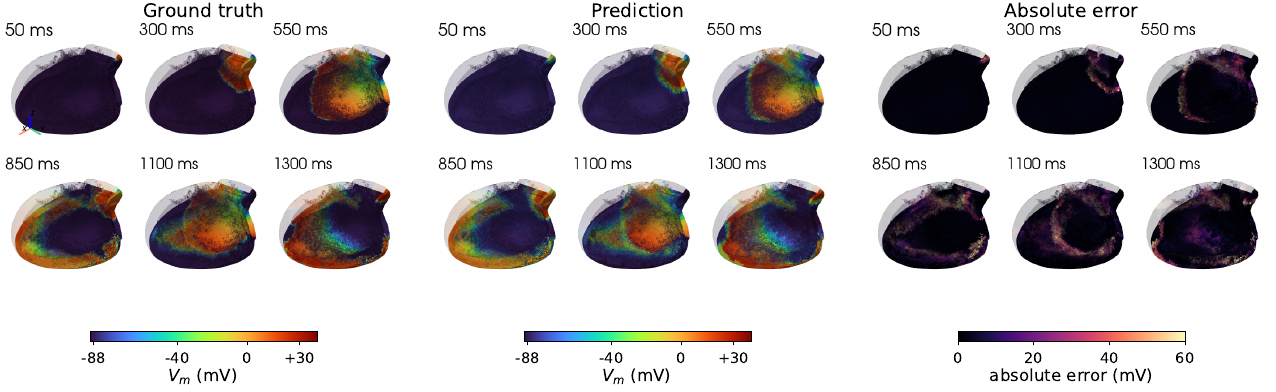}
  \caption{\small \textbf{GAP$^2$ODE prediction on a held-out scar sample.} Ground truth, prediction, and absolute error on the native mesh at six times spanning the paced sequence. Voltage panels share the $[-90, 40]\,\mathrm{mV}$ scale, error panels $[0, 60]\,\mathrm{mV}$. Error concentrates in a thin band at the moving wavefront while amplitude and morphology are preserved; the scar core, inactive and excluded from the simulated domain, appears as a cavity that the wave circumnavigates.}
  \label{fig:model-prediction}
\end{figure}

\begin{figure}[t]
  \centering
  \includegraphics[width=0.9\textwidth]{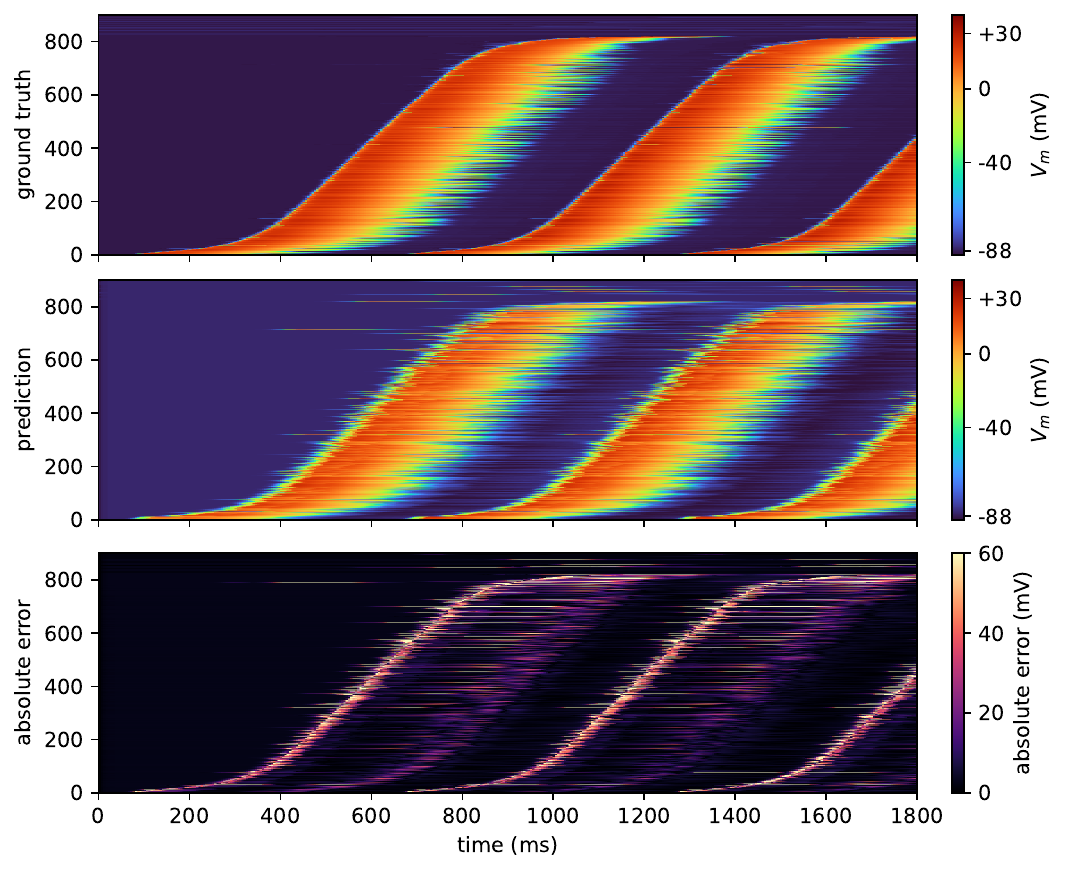}
  \caption{\small \textbf{Full-trajectory view of the same prediction.}
  Ground truth, prediction, and absolute error over all three paced beats, nodes ordered by first-beat activation time. The prediction preserves the propagation pattern across beats; error is confined to a narrow band tracking the wavefront.}
  \label{fig:model-prediction-heatmap}
\end{figure}

The stateful graph ODE achieves the lowest full-trajectory MSE and MAE, active-myocardium MAE, depolarisation-amplitude error, and LAT error in Table~\ref{tab:vm-benchmark}. Its MAE is $10.21\pm4.62\,\mathrm{mV}$, compared with $15.55\pm5.09\,\mathrm{mV}$ for HGT, the strongest one-shot model. Its depolarisation-amplitude error is $6.98\pm3.55\,\mathrm{mV}$ and LAT error $103.2\pm124.5\,\mathrm{ms}$, supporting explicit state evolution and sample-specific diffusion for propagation and amplitude preservation. Figure~\ref{fig:model-prediction} shows the prediction on a held-out scar sample: error concentrates in a thin band at the moving wavefront while amplitude and morphology are preserved, and Figure~\ref{fig:model-prediction-heatmap} shows the corresponding full-trajectory view, with the propagation pattern preserved across all three beats.

The gain does not extend to recovery: HGT-Lag retains the lowest APD90 and RT errors, while the graph ODE obtains $456.4\pm135.4$ and $422.3\pm174.4\,\mathrm{ms}$. Endpoint-dependent rankings show that voltage reconstruction, activation timing, and recovery morphology are distinct challenges. The graph ODE also changes from one shot to autoregressive; therefore, it reads the stimulation at every step and maintains the recovery state. This comparison tests state evolution and the biophysical features jointly rather than architecture alone. Statistical distributions of the performance of the models on the test samples are reported in the Extended results section at Figure~\ref{fig:model-capability-distributions}. Examples of predictions are reported in the extended section~\ref{sec:supp-prediction-examples}.

\begin{table}[htbp]
% \floatconts{tab:vm-benchmark}%
% {Full-trajectory benchmark under a maximum budget of 4,000 optimiser updates. All metrics are computed over the complete 361-frame ($0$--$1800\,\mathrm{ms}$) repaired-target test trajectories. MSE: mean squared error; MAE: mean absolute error; A-MAE: active-myocardium MAE; AR: depolarisation-amplitude ratio; DA-MAE: depolarisation-amplitude MAE; LAT: local-activation-time MAE; APD90: action-potential-duration-at-90\%-repolarisation MAE; RT: repolarisation-time MAE. Values are test mean $\pm$ sample standard deviation over the same 73 held-out simulations. $^{\dagger}$Trained with the same objective, converging at ${\sim}2{,}600$ optimiser updates, within the 4{,}000-update budget.}
\caption{Full-trajectory benchmark under a maximum budget of 4,000 optimiser updates. All metrics are computed over the complete 361-frame ($0$--$1800\,\mathrm{ms}$) repaired-target test trajectories. MSE: mean squared error; MAE: mean absolute error; A-MAE: active-myocardium MAE; AR: depolarisation-amplitude ratio; DA-MAE: depolarisation-amplitude MAE; LAT: local-activation-time MAE; APD90: action-potential-duration-at-90\%-repolarisation MAE; RT: repolarisation-time MAE. Values are test mean $\pm$ sample standard deviation over the same 73 held-out simulations. $^{\dagger}$Trained with the same objective, converging at ${\sim}2{,}600$ optimiser updates, within the 4{,}000-update budget.}
\label{tab:vm-benchmark}
\centering
{\scriptsize
\setlength{\tabcolsep}{2pt}
\resizebox{\linewidth}{!}{%
\begin{tabular}{@{}lrrrrrrrr@{}}\toprule
Model & MSE & MAE & A-MAE & AR & DA-MAE & LAT & APD90 & RT \\
 & (mV$^2$) & (mV) & (mV) &  & (mV) & (ms) & (ms) & (ms) \\\midrule
 GATv2 (4-layer) & $1762.5\pm353.7$ & $35.33\pm4.29$ & $38.08\pm3.60$ & $0.48\pm0.12$ & $63.18\pm9.39$ & $1238.0\pm242.6$ &
  $583.4\pm103.6$ & $676.4\pm231.2$ \\
  GHR & $870.1\pm463.0$ & $20.31\pm5.65$ & $21.70\pm7.49$ & $0.82\pm0.15$ & $28.92\pm8.17$ & $282.1\pm173.2$ &
  $463.4\pm123.7$ & $450.4\pm183.7$ \\
  GINO & $921.5\pm752.2$ & $20.78\pm8.27$ & $21.36\pm10.27$ & $0.89\pm0.13$ & $23.85\pm8.99$ & $184.7\pm178.4$ &
  $485.7\pm153.8$ & $448.2\pm189.2$ \\
  GWNO & $913.2\pm713.5$ & $20.54\pm8.26$ & $21.10\pm10.31$ & $0.89\pm0.13$ & $22.45\pm8.51$ & $173.8\pm171.5$ &
  $434.9\pm116.3$ & $413.5\pm163.5$ \\
  HGT & $612.6\pm433.6$ & $15.55\pm5.09$ & $15.98\pm6.34$ & $0.94\pm0.13$ & $18.77\pm10.50$ & $173.2\pm177.7$ &
  $434.5\pm104.6$ & $408.7\pm147.1$ \\
  HGT-Lag & $761.2\pm643.6$ & $16.86\pm7.61$ & $17.61\pm9.23$ & $0.94\pm0.17$ & $19.46\pm13.69$ & $237.5\pm261.3$ &
  $375.9\pm130.0$ & $291.2\pm115.4$ \\
  GAP$^2$ODE$^{\dagger}$ & $424.8\pm406.6$ & $10.21\pm4.62$ & $10.44\pm5.97$ & $0.95\pm0.04$ & $6.98\pm3.55$ & $103.2\pm124.5$ &
  $456.4\pm135.4$ & $422.3\pm174.4$ \\
\bottomrule\end{tabular}%
}%
}
\end{table}

\section{Discussion and Conclusion}
GraphEP addresses the gap between producing a mechanistic simulation and turning it into reusable SciML data. A voltage trajectory is meaningful only in relation to the anatomy, material properties, calibration and intervention that produced it. By retaining this chain through graph conversion and deterministic release, GraphEP makes provenance part of the data representation rather than an external account of dataset construction.

The benchmark shows how model structure interacts with this representation. Communication beyond local neighbourhoods improves one-shot prediction, while physiological biases provide complementary benefits. The graph ODE improves voltage, depolarisation amplitude and activation timing by evolving recovery state with a sample-specific diffusion operator; HGT-Lag remains more accurate on recovery endpoints by separating propagation timing from waveform morphology. Neither is uniformly superior, suggesting that propagation and repolarisation reward different physical biases.

Stateful surrogates also require a broader data contract than immutable trajectories alone: they need initial electrical state, time-resolved interventions and a reproducible physical operator. GraphEP supplies simulation-derived quantities as biophysical features, while recovery and event history remain internal to the rollout. This distinction prevents target-derived information from being mistaken for model state and allows the same release to support both one-shot and autoregressive learning.

GraphEP remains a controlled simulation testbed rather than a substitute for clinical data. Its split evaluates new conditions on 20 known anatomies and is not disjoint, so performance on unseen anatomies is yet untested. Scar patterns are procedurally generated and have not been validated against clinical scar segmentations or histopathology. The reported model rankings therefore describe this synthetic simulation distribution and may change for clinically derived scars or unseen anatomies. The current representation can hide non-resting boundary state, mix activation events across beats or sparse fast-conduction tissue during coarsening. Learned operator corrections may likewise restore flux across scar boundaries unless constrained by tissue semantics. These limitations motivate anatomy-disjoint evaluation, event-local labels, rare-tissue preservation and stricter operator enrichment. Within these bounds, GraphEP provides a reproducible setting for studying data design and model inductive bias together, with future extensions toward broader physiology, uncertainty and measured cardiac signals. Availability of the data is supported at the Zenodo archive v1.2 \href{https://doi.org/10.5281/zenodo.24085325}{https://doi.org/10.5281/zenodo.24085325}.

\begin{ack}
James Rowbottom gratefully acknowledges support from the Horizon Europe projects TROPHY (ultrafast holographic FTIR microscopy, 2022--2026) and CHARM (coherent Raman imaging for precision medicine). The authors acknowledge support from the National Science Foundation AI Institute in Dynamic Systems (grant no.~2624398). Matteo Baldan was supported by the National Center for Gene Therapy and Drugs Based on RNA Technology, funded by the European Union---Next-Generation EU under Italy's National Recovery and Resilience Plan (NRRP), Mission~4, Component~2, Investment~1.4 (Project CN00000041, CUP B93D21010860004, Spoke~2 ``Cancer''). The authors thank TaccLab and DAWN for providing computational resources.
\end{ack}

\bibliographystyle{plain}
\bibliography{biblio}

\newpage
\appendix

\section{Supplementary Data and Methods}
\subsection{Simulation pipeline: methods and configuration}

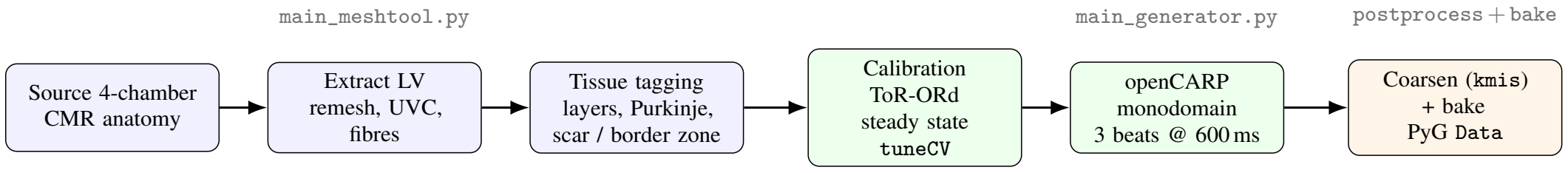
\begin{figure}[htbp]
\centering
\resizebox{\linewidth}{!}{%
\begin{tikzpicture}[
  font=\footnotesize,
  box/.style={draw, rounded corners, align=center, inner sep=4pt, minimum height=11mm, text width=24mm},
  mtool/.style={box, fill=blue!6},
  gen/.style={box, fill=green!7},
  bake/.style={box, fill=orange!9},
  arr/.style={-{Latex[length=2.4mm]}, thick},
    lab/.style={font=\footnotesize, text=black!55}
]
\node[mtool] (src) {Source 4-chamber\\CMR anatomy};
\node[mtool, right=6mm of src] (mesh) {Extract LV\\remesh, UVC,\\fibres};
\node[mtool, right=6mm of mesh] (tag) {Tissue tagging\\layers, Purkinje,\\scar / border zone};
\node[gen, right=8mm of tag] (cal) {Calibration\\ToR-ORd steady state\\\texttt{tuneCV}};
\node[gen, right=6mm of cal] (sim) {openCARP\\monodomain\\3 beats @ 600\,ms};
\node[bake, right=8mm of sim] (bake) {Coarsen (\texttt{kmis})\\ + bake\\PyG \texttt{Data}};
\foreach \a/\b in {src/mesh, mesh/tag, tag/cal, cal/sim, sim/bake} \draw[arr] (\a) -- (\b);
\node[lab, above=3mm of mesh] {\texttt{main\_meshtool.py}};
\node[lab, above=3mm of sim] {\texttt{main\_generator.py}};
\node[lab, above=3mm of bake] {\texttt{postprocess}\,$+$\,\texttt{bake}};
\end{tikzpicture}}
\caption{GraphEP data-generation pipeline. Mesh preparation (blue) and EP
generation (green) are each driven by a single JSON-configured entry point; a
postprocessing/baking stage (orange) coarsens the finite-element graph and
writes self-contained PyTorch Geometric samples.}
\label{fig:pipeline-overview}
\end{figure}

GraphEP is produced by a fully scripted, four-stage pipeline (Figure~\ref{fig:pipeline-overview}): mesh preparation turns a source four-chamber anatomy into a tagged, fibre-annotated openCARP mesh; physiological calibration brings the ionic model to periodic steady state and fits conductivities to prescribed conduction velocities; openCARP solves the monodomain equations over three paced beats; and a postprocessing/baking stage that coarsens the finite-elment graph and prepare self-contained PyTorch Geometric samples. Coarsening is obviously optional but important to let model training scale. JSON-configured scripts drive the mesh and simulation, respectively. Every stage writes an auditable intermediate.

\subsection{Mesh handling and augmentation}
\label{sec:supp-mesh-handling}
Mesh preparation is recipe-driven and is executed by \texttt{interfaces/main\_meshtool.py}.  A recipe identifies the immutable source anatomy, the chambers to retain, the desired spatial resolution, the universal ventricular coordinate (UVC) transfer method, and an ordered list of tissue-tagging operations.  The pipeline has four auditable stages: 
\begin{enumerate}   
    \item \emph{Conversion.} The source VTK anatomy is converted to the openCARP \texttt{.pts/.elem/.lon} representation. Coordinates are normalized to $\mu$m, and the UVC fields are copied from point data or averaged from cell data to mesh vertices when necessary.   
    \item \emph{Chamber extraction.} Source element tags select the LV alone, both ventricles, or all four chambers. The extracted boundary is smoothed when requested, while the original source tags are retained as immutable provenance.
    \item \emph{Resolution and coordinates.} The tetrahedral mesh is uniformly or curvature-adaptively resampled subject to the recipe's target and minimum/maximum edge scales.  UVC fields are then recomputed or interpolated onto the new vertices; the fibre and sheet directions remain associated with the resampled elements.
    \item \emph{Semantic tagging.} Ordered tagging methods assign atrial and ventricular regions, UVC-defined endocardial/mid-myocardial/epicardial layers, Purkinje tissue, and optional pathology.  The final integer mapping is written to \texttt{mesh.tags.json}, so downstream simulation and graph baking consume labels rather than campaign-specific tag numbers.
\end{enumerate}
Each stage writes a separate intermediate and is skipped only when its expected mesh and metadata already exist.  This makes extraction, remeshing, coordinate transfer, and biological labelling independently inspectable.

\paragraph{Resolution and uvc coordinates}
The resolution is rescaled accordingly to three configurable parameters: the expected median edge length, which would represent the median cardiac fibre length or the media tetrahedra face edge length; and a minimum and maximum value of such range. Then, a \texttt{resample} tool from \texttt{meshtool} is iteratively called to aggregate or cut edges of adjacent tetrahedra to converge to the desired choice.
Universal Ventricular Coordinates instead (UVC) include four scalar fields as follows:
\begin{itemize}
    \item transmural $\rho$: ranges between $[0,1]$ and represents the normalised depth through the ventricular wall.
    \item rotational $\psi$: ranges between $[-\pi,\pi]$ and represents the ventricular long axis, distinguishing the septal, anterior, lateral and posterior position at a fixed depth and height.
    \item apico-basal $z$: ranges in $[0,1]$ and runs along the long axis from the apex at $z=0$ to the base $z=1$.
    \item ventricle identifier $v$: represent the identifier of the left or right ventricle.
\end{itemize}
These coordinates can be provided or recomputed after remeshing by solving a Laplace problem with anatomically defined boundary conditions \cite{uvc}.

\paragraph{Semantic tagging}
Healthy tissue is defined by using intervals of the transmural coordinate that ranges between $[0,1]$ where 0 lands in the center of the geometry and 1 on the extent of the bounding box. So we can use thresholds on this coordinate to divide the volume in three sections from internal to external, i.e., endocardium; mesocardium and pericardium. 

Pathological regions are incorporated by modifying the electrophysiological properties after geometric tagging.  Core scar receives no ionic model and is assigned a uniform isotropic bath conductivity of $10^{-4}$ S/m, which keeps the monodomain operator nonsingular while making the region effectively non-conductive.  Border zones are KD-tree distance shells around core-scar elements \cite{maneewongvatana1999analysis}. Border-zone tissue retain an active Tomek model but use 10\% lower transverse conductivity.  Random-walker scars are formed on the element adjacency graph by multiple walkers biased toward elements not visited by earlier walks. The union of visited elements defines the core, after which the distance shell is assigned. The walkers, as summarised in \ref{alg:long-wander}, start from a surface and are biased towards new nodes that they have not seen yet. They are more likely to remain on the surface but they are allowed with a lower probability to dive into the volume.

 \begin{algorithm}[htbp]
\caption{Biased random-walk generation of scar tissue}
\label{alg:long-wander}
\begin{algorithmic}[1]
\Require element adjacency $\mathrm{adj}$, endocardial surface $\mathcal{B}$,
         seed elements $S$
\Ensure scar core $V$, border zone $R$

\State \textbf{Set} number of walkers $W$, walk length $L$,
       novelty probability $\alpha$, surface preference $p_{\mathrm{surf}}$,
       border-zone thickness $\tau$
\State $V \gets S$
\Comment{Initialise the seen set of nodes}

\ForAll{$s \in S$}
    \For{$w \gets 1$ \textbf{to} $W$}
        \State $c \gets s$
        \For{$t \gets 1$ \textbf{to} $L$}
            \State $C \gets \mathrm{adj}(c)$

            \If{$\mathrm{rand}() < p_{\mathrm{surf}}$ 
                \textbf{and} $C \cap \mathcal{B} \neq \emptyset$}
                \State $C \gets C \cap \mathcal{B}$
                \Comment{prefer endocardial propagation}
            \EndIf

            \If{$\mathrm{rand}() < \alpha$
                \textbf{and} $C \setminus V \neq \emptyset$}
                \State $C \gets C \setminus V$
                \Comment{prefer unexplored elements}
            \EndIf

            \State $c \sim \mathrm{Unif}(C)$
            \State $V \gets V \cup \{c\}$
        \EndFor
    \EndFor
\EndFor

\State $R \gets
\{e \notin V :
\operatorname{dist}(e,V) \leq \tau\}$
\Comment{border zone surrounding the scar core}

\State \Return $V,R$
\end{algorithmic}
\end{algorithm}

Purkinje network fibres, instead,  are generated using a fractal tree set of rules. The LV and RV trees are generated independently on their endocardial surfaces. For each chamber, boundary triangles are first extracted from the tetrahedral volume.  Triangle normals distinguish the inward-facing cavity surface, and a principal-axis decomposition identifies the apical and basal directions.  The configured root is either the apex, a reproducibly sampled apical face, a reproducibly sampled endocardial face, or an explicit vertex.  The initial growth direction follows a surface edge toward the base where possible and is otherwise obtained by projecting the apex--base vector onto the local tangent plane.

The fractal-tree generator grows a surface-constrained trunk and recursively bifurcating branches and is implemented using the source methodology of Costabal et al. \cite{costabal2016generating}.  Its principal controls are iteration count, initial and daughter-branch lengths, discretization length, branch angle, repulsion from existing branches, and optional fascicles.  In the implemented adaptive mode, the discretization length is derived from the harmonic mean of endocardial triangle-edge lengths; the trunk, branches, and fascicles scale with chamber long-axis length. Coverage is measured as the fraction of endocardial vertices within a prescribed Euclidean radius of a tree node.  If the target is missed, bounded retries increase the iteration count and shorten daughter branches; the highest-coverage realisation is retained. For volumetric Purkinje representations, a KD-tree tags tetrahedra whose centroids fall within the configured distance of a branch. The authors suggest monitoring the generation of the purkinje tree based on the spatial resolution of the mesh.

\subsection{Physiological calibration}

To ensure that the anisotropy of the diffusion tensor produces physiologically accurate propagation behaviour at the chosen spatial resolution, the initial seed conductivities are refined using openCARP's tuneCV tool \cite{gsell2024forcepss}. In that tool, for each distinct tissue region, calibration is performed on a 2-cm 1-D monodomain cable. Such cable is initialized from the periodic cellular steady state, derived from a previous stage where a basic cycle of length 1000ms is run. The purpose of this BCL is to ensure that the membrane is physiologically polarized before the first action potential is launched, making the measured CV more representative of tissue in a steady pacing state. Extracellular conductivity is fixed, while intracellular conductivity is iteratively adjusted along a single principal axis (fibre, sheet, or normal) in each run, until the measured CV matches the reference (within a tolerance).

Given that a fixed stimulus current often fails in high-conductivity tissues or causes non-physical voltage spikes in low-conductivity tissues, we use a square-root scaling of the stimulus strength based on the seed intracellular conductivity, mimicking the dependence of the cable length constant and guaranteeing supra-threshold depolarization across all tissue types without PDE divergence. This follows suggestions from the authors of ForCEPPs \cite{gsell2024forcepss}.

Pathological regions are incorporated through the modification of electrophysiological properties, whereby scar tissue is not assigned an ionic model since it is not electrically active, while border zone regions exhibit reduced (10\% less than healthy) transverse conductivity to emulate fibrosis-induced slowing. Note that the scar region is given a uniform isotropic bath conductivity (of 1×$10^{-4}$ S/m), in order to maintain monodomain operator non-singularity while considering the region as non-conductive. 
Border zones are defined as a distance-based shell (computed with KD-tree) around core scar elements. 

\begin{figure}
    \centering
    \includegraphics[width=0.75\linewidth]{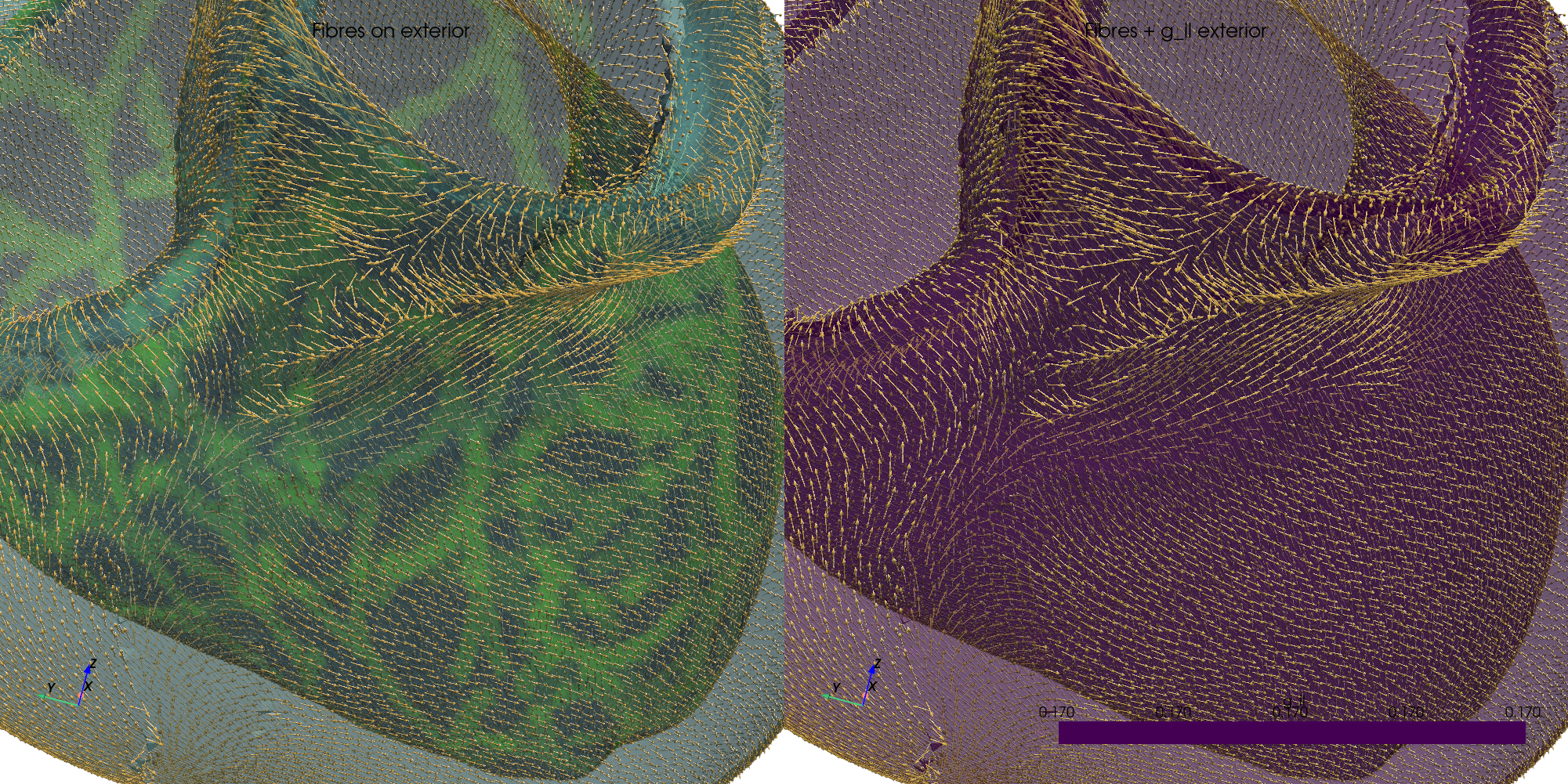}
    \caption{Local fibre directions in a tagged ventricular mesh. The orthotropic conductivity tensor is aligned with the fibre, sheet and sheet-normal frame, producing anisotropic propagation through the myocardium.}
    \label{fig:fiberdirection-example}
\end{figure}

In Fig ~\ref{fig:fiberdirection-example} we report a zoom of a real heart volumetric mesh, with tagged regional transmural endocardium, mesocardium, and epicardium regions, Purkinje network, and fibre directions.

\subsection{openCARP monodomain solver}
Electrical propagation is modelled by coupling the monodomain equation to the ToR-ORd ventricular-cell dynamics~\cite{tomektorord}:
\begin{equation}
\begin{aligned}
\beta C_m \frac{\partial V_m}{\partial t}
&=\nabla\!\cdot\!\left(\boldsymbol\sigma_m\nabla V_m\right)
-\beta I_{\mathrm{ion}}^{\mathrm{ToR\text{-}ORd}}(V_m,\mathbf q)
+I_{\mathrm{stim}},\\
\frac{\mathrm d\mathbf q}{\mathrm dt}
&=\mathbf F_{\mathrm{ToR\text{-}ORd}}(V_m,\mathbf q),
\end{aligned}
\label{eq:monodomain-torord}
\end{equation}
where $C_m$ is membrane capacitance, $\beta$ is the membrane surface-to-volume ratio, $\boldsymbol\sigma_m$ is the anisotropic conductivity tensor, $\mathbf q$ denotes the ToR-ORd cellular state variables and $I_{\mathrm{stim}}$ is the applied stimulus. Ventricular ionic dynamics are described by the Tomek--Rodriguez modification of the O'Hara--Rudy model (ToR-ORd). ToR-ORd is a biophysical human ventricular cardiomyocyte model in which the total ionic current $I_{\mathrm{ion}}$ results from the balance of transmembrane sodium, calcium and potassium currents together with intracellular calcium cycling. Its gating variables and ionic concentrations form the local state $\mathbf q$, whose evolution generates the rapid action-potential upstroke, plateau and repolarisation phases. Endocardial, mid-myocardial, and epicardial parameterizations reproduce regional differences in ventricular action-potential morphology. In the tissue simulation, ToR-ORd supplies the node-local reaction term, while openCARP couples neighbouring cells through the finite-element discretisation of the anisotropic monodomain operator.

\subsection{Conversion of EP into PyTorch Geometric objects}
The baking stage converts each finite-element solution into the graph tensors defined in Equation~\eqref{eq:dataspace}. Native vertices are assigned to retained KMIS nodes through the same partition used to aggregate extensive node volume. Geometry, tissue, stimulation and voltage fields are pooled according to their semantics: continuous intensive quantities are averaged, extensive volume is summed, stimulus masks use logical OR and categorical tissue labels are inherited from representative nodes. The frozen PyG object remains the immutable release unit.

\subsection{Offline operator and distance-field enrichment}
\label{sec:supp-enrichment}
Some models require fields reproducible from archived meshes and plans but absent from the original 17-channel tensor. These additional features include; signed distance fields that encode proximity to the scar interface and core/border-zone region. For physical diffusion, the native $P_1$ stiffness matrix is assembled from
\begin{equation}
\boldsymbol\sigma_e=g_{f,e}\mathbf f_e\mathbf f_e^\top+g_{s,e}\mathbf s_e\mathbf s_e^\top+g_{n,e}\mathbf n_e\mathbf n_e^\top,
\end{equation}
and coarsened with the bake partition,
\begin{equation}
K_c=P^\top KP,\qquad M_c=P^\top MP.
\end{equation}
Non-zero coarse edge coefficients and lumped masses are attached as \texttt{fem\_edge\_stiffness} and \texttt{fem\_node\_mass}. The transformation records its source mesh, plan and enrichment version. The released audit must define the stiffness sign convention and the handling of negative off-diagonal terms.

Additional biophysical features are distinct from the online state. Time since depolarisation, preceding diastolic interval, beat phase, and voltage history are updated from predicted states during free rollout; they are not target-derived features at evaluation time.

\section{Models}
\subsection{Autoregressive state-evolution models}
\label{sec:supp-graph-ode}

\subsubsection{State and reaction dynamics}
Each node carries $z_i=(u_i,w_i,h_i)$, where $u_i$ is normalised excitation, $w_i$ is recovery, and $h_i$ is optional latent memory. The physical core uses dynamics like the  Aliev--Panfilov model,
\begin{align}
\dot u_i &= r_i[-k u_i(u_i-a_i)(u_i-1)-u_iw_i]+I_i^{\mathrm{diff}}+s_i(t)q(w_i),\\
\dot w_i &= \varepsilon_i(u_i)[ku_i(1+a_i-u_i)-w_i],
\end{align}
with $k=8$ and refractory gate $q(w)=\sigma(8(w^*-w))$, $w^*=0.35$. Learned heads modulate bounded positive reaction rates and edge-conductance gains. The authoritative configuration must state whether any additive state residual is enabled; rate modulation preserves the equilibrium structure more directly than an unconstrained additive update.

\subsubsection{Coarsened FEM diffusion}
Let $L_c$ denote the coarse stiffness/Laplacian under the stored sign convention and $M_c$ its lumped mass. Diffusion is applied as a conservative flux,
\begin{equation}
I^{\mathrm{diff}}=-\delta M_c^{-1}L_c(G\odot u),
\end{equation}
where $G$ is a bounded learned edge correction conditioned on edge geometry and tissue. This correction closes the operator error introduced by coarsening but must be audited near scar boundaries.

\subsubsection{Online physiological state and observation map}
Latched predicted threshold crossings maintain the time since the last upstroke and the preceding diastolic interval. Beat phase and short predicted-voltage history can condition the parameter and readout heads. The observation map
\begin{equation}
\widehat V_i(t)=V_{\mathrm{rest}}+A[u_i(t)+\psi_\theta(u_i,w_i,h_i,X_i)]
\end{equation}
uses bounded $\psi_\theta$ to reproduce waveform morphology without feeding that correction back into the dynamical state.

\subsubsection{Integration and training}
Reaction and stimulation are integrated explicitly with RK2, while stiff diffusion uses implicit Euler with a fixed-iteration, Jacobi-preconditioned conjugate-gradient solve. Rolling the transition over all frame times produces the 361-frame prediction. The archived run record must identify initial-state construction, objective, optimiser updates, solver, substeps, implicit stride, Conjugate Gradients iterations, and enrichment version. Target states or target-derived event timers must not enter headline free-rollout evaluation.

\subsection{Local receptive field}
\subsubsection{GATv2}
GATv2 is a message-passing graph neural network in which node features are first embedded using a two-layer MLP $(17 \to d)$, followed by L stacked GATv2 layers with multi-head attention \cite{brody2021attentive}. Multi-head outputs are averaged to preserve a constant hidden dimensionality across layers. A final two-layer MLP projects the latent node representations to the desired output dimension (361 for $V_m$ or 1 for LAT prediction). Normalised edge length, effective conductivity projected along the edge direction %{, and normalized tissue labels for the source and destination nodes }
are encoded as edge features. These are then incorporated into the attention mechanism to represent anisotropic cardiac conduction. The model uses hidden width d=64, attention heads=4, 3 GATv2Conv layers, and no dropout.

\subsection{Multiscale receptive fields}
% Supplementary: mathematical description of HGT / HGT-Lag.
% Matches src/heartmodeling/machinelearning/models/multiscale/hierarch_graph_transf.py
% and .../hgt_lag.py.  Uses only amsmath.
\subsubsection{HGT and HGT-Lag: layer definitions}
\label{sec:supp-hgt}

\paragraph{Notation.}
A sample has $N$ nodes with positions $p_i\in\mathbb{R}^3$ and features $x_i\in\mathbb{R}^{F}$, and $T$ output frames at times $t_1,\dots,t_T$ in ms. Hidden vectors have width $d$; attention uses $H$ heads of width $d_h=d/H$. The encoder has $L$ levels with grid resolutions $\Gamma_1>\dots>\Gamma_L$. All blocks below act on one graph at a time; in a batch they are applied independently per graph.

\begin{figure}[htbp]
    \centering
    \includegraphics[width=1\linewidth]{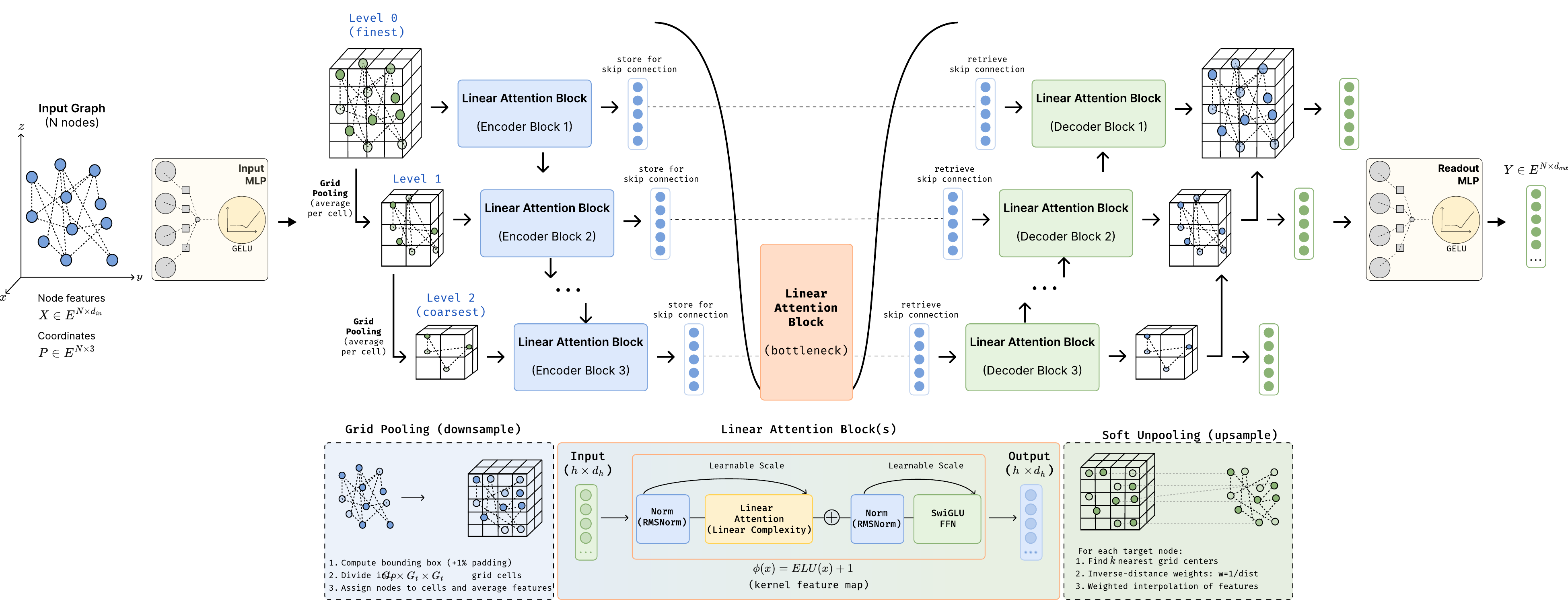}
    \caption{Architecture outline of HGT. }
    \label{fig:hgt_arch}
\end{figure}

\paragraph{Input embedding.}
Node features are the 17 baked channels concatenated with a 16-dimensional encoding of the node's stimulus time series, produced by a small pointwise network. They are lifted once,
\begin{equation}
h_i^{(0)}=\mathrm{GELU}\!\left(\mathrm{MLP}_{\mathrm{in}}(x_i)\right)\in\mathbb{R}^{d}.
\end{equation}

\paragraph{Grid pooling.}
At level $\ell$ the bounding box of the graph is divided into  $\Gamma_\ell^3$ equal cells. Let $c(i)$ be the cell containing $p_i$ and $S_c=\{i: c(i)=c\}$ the nodes falling in cell $c$. Only non-empty cells are kept, and each becomes one token with the mean feature and the mean position of its nodes,
\begin{equation}
h_c^{(\ell)}=\frac{1}{|S_c|}\sum_{i\in S_c} h_i^{(\ell-1)},
\qquad
p_c^{(\ell)}=\frac{1}{|S_c|}\sum_{i\in S_c} p_i .
\end{equation}
Pooling is, therefore, a pure averaging operator with no learnable parameters, and the number of tokens drops from $N$ to the number of occupied cells.

\paragraph{Linear-attention block.}
Let $\{\zeta_c\}_{c=1}^{n}$ be the tokens of one level. Write
$\mathrm{RMS}(\zeta)=\zeta\,/\sqrt{\overline{\zeta^2}+\varepsilon}$ for the element-wise root-mean-square normalisation with a learnable gain, and let $\phi(u)=\mathrm{ELU}(u)+1$ be the positive feature map. With $u_c=\mathrm{RMS}(\zeta_c)$ the projections are
\begin{equation}
q_c=\phi(W_q u_c),\qquad k_c=\phi(W_k u_c),\qquad v_c=W_v u_c ,
\end{equation}
each reshaped into $H$ heads. For every head, the attention output is
\begin{equation}
a_c=\frac{q_c^{\top}\sum_{j=1}^{n} k_j v_j^{\top}}{q_c^{\top}\sum_{j=1}^{n} k_j},
\end{equation}
so the two sums are accumulated once per level, and the cost is $O(n\,d\,d_h)$ rather than $O(n^2 d)$: every token still sees every other token at that scale. The block updates its input twice, with learnable scalars $\gamma_{\mathrm{attn}},\gamma_{\mathrm{ffn}}$ initialised at zero,
\begin{align}
\zeta_c &\leftarrow \zeta_c+\gamma_{\mathrm{attn}}\,W_o\,a_c ,\\
\zeta_c &\leftarrow \zeta_c+\gamma_{\mathrm{ffn}}\,
\mathrm{FFN}\!\left(\mathrm{RMS}(\zeta_c)\right),
\qquad
\mathrm{FFN}(u)=W_2\left(\mathrm{SiLU}(W_g u)\odot W_u u\right),
\end{align}
where $\odot$ is the element-wise product. Both updates are residual, so a block starts as the identity and learns how much to add.

\paragraph{Unpooling.}
To return from the tokens of level $\ell$ to the $N_{\ell-1}$ nodes of the finer level, each target node $i$ takes the $k=3$ nearest pooled centroids $\mathcal{N}(i)$ and interpolates with inverse-distance weights, 
\begin{equation}
\tilde h_i=\sum_{c\in\mathcal{N}(i)} w_{ic}\,h_c ,
\qquad
w_{ic}=\frac{\lVert p_i-p_c\rVert^{-1}}{\sum_{c'\in\mathcal{N}(i)}\lVert p_i-p_{c'}\rVert^{-1}} .
\end{equation}

\paragraph{Encoder, bottleneck and decoder.}
The encoder stores a skip copy, pools, and applies one block per level; the bottleneck applies one block at the coarsest level; the decoder unpools, concatenates the stored skip, projects back to width $d$ and applies one block:
\begin{align}
s^{\mathrm{skip},(\ell)}&=h^{(\ell-1)},\qquad
h^{(\ell)}=\mathrm{Block}_\ell\!\left(\mathrm{Pool}_{\Gamma_\ell}(h^{(\ell-1)})\right),
&&\ell=1,\dots,L,\\
h&\leftarrow \mathrm{Block}_{\mathrm{bott}}(h^{(L)}),\\
h&\leftarrow \mathrm{GELU}\!\left(\mathrm{Block}^{\mathrm{dec}}_{\ell}\!\left(
W_{\mathrm{dec}}^{\ell}\left[\,\mathrm{Unpool}(h)\;;\; s^{\mathrm{skip},(\ell)}\,\right]\right)\right),
&&\ell=L,\dots,1 .
\end{align}

The result is one vector $h_i\in\mathbb{R}^{d}$ per original node.

\paragraph{Decoder A: HGT.}
The trajectory is read out pointwise, all frames at once,
\begin{equation}
\hat V_i=\mathrm{MLP}_{\mathrm{out}}(h_i)\in\mathbb{R}^{T},
\end{equation}
so time is an output dimension, and nothing constrains the predicted waveform.

\paragraph{Decoder B: HGT-Lag.}
A linear head produces three scalars per node, and a second head produces the timing residual,
\begin{equation}
(A_i,\;b_i,\;\lambda_i)=W_{\mathrm{lag}}h_i,
\qquad
v_i=\mathrm{softplus}(\lambda_i)+10^{-3},
\qquad
\xi_i=\tanh\!\left(w_\xi^{\top}h_i\right).
\end{equation}
With $d_i$ the geodesic distance from node $i$ to the stimulation site and $\Delta_{\max}$ the residual bound, the activation time is
\begin{equation}
\eta_i=\max\!\left(0,\;\frac{d_i}{v_i}+\Delta_{\max}\,\xi_i\right).
\label{eq:supp-hgt-tau}
\end{equation}
The shared template is an analytic upstroke of learnable width $w$ multiplied by a learnable tail $g_{\mathrm{tail}}$, both evaluated at the physical offset $\delta=t-\eta_i$,
\begin{equation}
g(\delta)=\frac{1}{1+e^{-\delta/w}}\;g_{\mathrm{tail}}(\delta),
\qquad
w=\mathrm{softplus}(\omega)+\gamma_{\min},
\end{equation}
where $g_{\mathrm{tail}}$ is stored on a fixed 5 ms grid and read by linear interpolation. The trajectory is the shifted, scaled template returned to millivolts,
\begin{equation}
\hat V_i(t)=V_{\mathrm{scale}}\left[A_i\,g(t-\eta_i)+b_i\right]+V_{\mathrm{rest}} .
\end{equation}
Since the upstroke factor equals $1/2$ at $\delta=0$, $\eta_i$ is the predicted activation time itself and can be supervised directly against the activation times. Before training, the amplitude bias, the upstroke width and the conduction-speed bias are initialized from robust statistics of the training split -- the median peak amplitude, the median maximum upstroke rate and the median ratio of distance to activation time -- rather than by averaging trajectories, which would blur the upstroke.

\begin{figure}[ht]
    \centering
    \includegraphics[width=\linewidth]{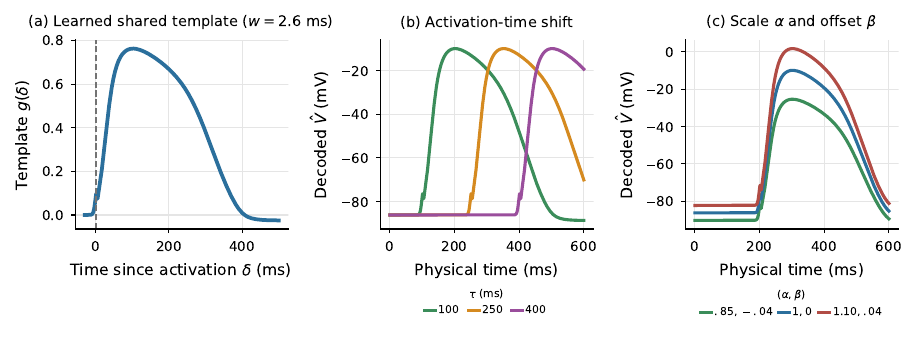}
    \caption{Behaviour of the HGT-Lag trajectory decoder. (a) Shared template learned by the selected propagation-aware checkpoint. (b) Controlled activation-time shifts of that template. (c) Controlled changes to its scale and offset at fixed activation time. The shifts and affine parameters in (b--c) are synthetic examples chosen to expose the decoder's operation; they are not benchmark predictions.}
    \label{fig:hgt-lag-decoder-behavior}
\end{figure}

Figure~\ref{fig:hgt-lag-decoder-behavior} illustrates the resulting separation of waveform morphology from activation timing. The shared learned template determines the common action-potential shape. The node-specific $\eta_i$ translates that shape in physical time, while $A_i$ and $b_i$ control its scale and offset. Thus the linear heads do not independently predict every displayed sample; they predict interpretable parameters through which the non-linear, time-continuous decoder constructs the trajectory.

\paragraph{Settings used in the benchmark.}
$d=128$, $H=4$, grid hierarchy $(\Gamma_1,\Gamma_2, \Gamma_3)=(16,8,4)$, $k=3$ interpolation neighbours, feed-forward width $\tfrac{4}{3}d$, no dropout, 16 stimulus channels. For HGT-Lag, $\Delta_{\max}=40$ ms, initial upstroke width $2$ ms with floor $w_{\min}=0.3$ ms, $V_{\mathrm{rest}}=-86.2$ mV and $V_{\mathrm{scale}}=100$ mV.

\paragraph{Decoder-component ablation.}
We isolate the temporal decoder while retaining the HGT backbone, repaired targets, frozen split, seed, loss and optimiser-update cap. Direct HGT regresses all frames independently through its output head; the template-only variant predicts activation lag directly and reconstructs voltage with the shared action-potential template; the propagation-aware variant augments the same template with the distance--speed prior and a bounded residual lag. Table~\ref{tab:hgt-lag-decoder-ablation} reports this controlled single-seed comparison.

\begin{table}[htbp]
\centering
\small
\caption{HGT-Lag decoder ablation on the frozen test split ($n=73$ simulations). The HGT backbone and training protocol are fixed; only the trajectory decoder changes. Values are test mean $\pm$ sample standard deviation for one matched training seed.}
\label{tab:hgt-lag-decoder-ablation}
\setlength{\tabcolsep}{3.5pt}
\begin{tabular}{@{}lccrrrr@{}}
\toprule
Decoder & AP temp. & Prop. prior & MAE & LAT & APD90 & RT \\
& & & (mV) & (ms) & (ms) & (ms) \\
\midrule
Direct HGT & no & no
  & $15.55\pm5.09$ & $173.2\pm177.7$ & $434.5\pm104.6$ & $408.7\pm147.1$ \\
Template only & yes & no
  & $20.69\pm 4.92$ & $310.9\pm205.2$ & $637.2\pm181.9$ & $489.2\pm242.1$ \\
Propagation aware & yes & yes
  & $16.86\pm7.61$ & $237.5\pm261.3$ & $375.9\pm130.0$ & $291.2\pm115.4$ \\
\bottomrule
\end{tabular}
\end{table}

Imposing the shared AP template without a propagation prior worsens every endpoint relative to Direct HGT: MAE increases by \$5.14\$ mV, while LAT, APD90 and RT errors increase by \$137.7\$, \$202.7\$ and \$80.5\$ ms, respectively. Adding the distance--speed propagation prior to that template improves every endpoint relative to Template only, reducing MAE by \$3.83\$ mV and LAT, APD90 and RT errors by \$73.4\$, \$261.3\$ and \$198.0\$ ms, respectively. The propagation-aware decoder nevertheless remains worse than Direct HGT on MAE and LAT, while improving APD90 and RT.

\subsubsection{Graph Hierarchical Recurrent (GHR)}
\begin{figure}[htbp]
        \centering
    \includegraphics[width=0.8\linewidth]{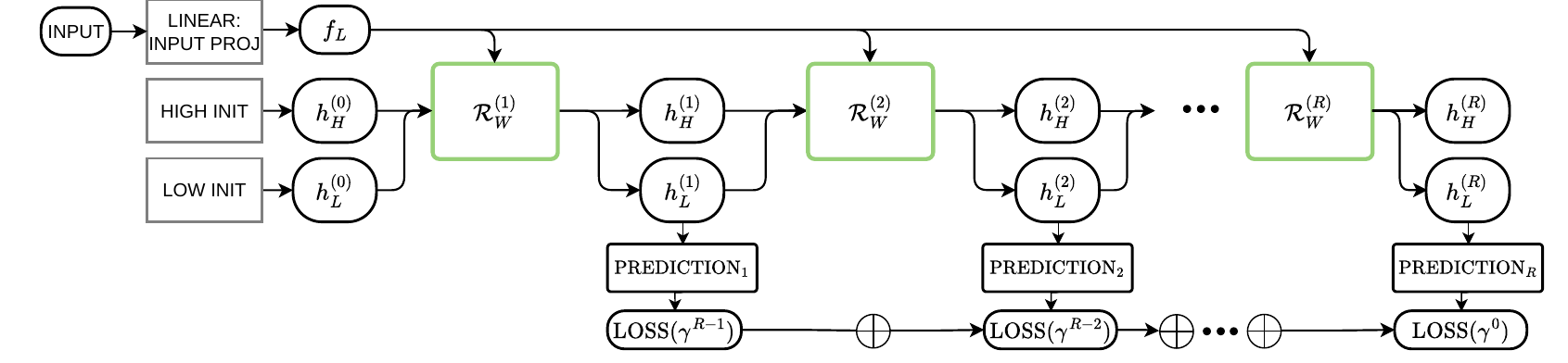}
    \caption{Architecture of GHR presented in the original contribution \cite{carotti2026graph}.}
    \label{fig:ghr}
\end{figure}

Graph Hierarchical Recurrent (GHR) neural network alternates message passing between fine- and coarse-scale graph representations, and its architecture is outlined in Figure \ref{fig:ghr} \cite{carotti2026graph}. At each forward pass, the input graph is coarsened (with Graclus clustering) to construct a two-level hierarchy. 
Through scatter reduction, node states are aggregated to coarse supernodes on the fine graph, updated using a coarse-level GINEConv, and then broadcasted back to the fine graph to guide subsequent fine-scale GINEConv updates. This bidirectional communication (fine-coarse) is repeated for several recurrent reasoning steps. This allows to propagate long-range information while also preserving local geometry structure. In this case, edge features are represented by relative (Cartesian) coordinates and Euclidean distance, all encoded separately for each graph level. Node embeddings from the final scale, are then mapped to predictions through a linear regression head. 
The configuration considered, includes 3 GINEConv layers per coarse iteration (for fine-level message passing) or 2 GINEConv iterations (for coarse level message passing), 5 hierarchical (recurrent) reasoning steps. Hidden dimension is 64, RMSNorm is used for normalization, and an MLP for activation. Total message-passing rounds are 30 (5x2x3). 

\subsection{Global receptive field}
\subsubsection{GiNO}
\begin{figure}[htbp]
    \centering
    \includegraphics[width=0.8\linewidth]{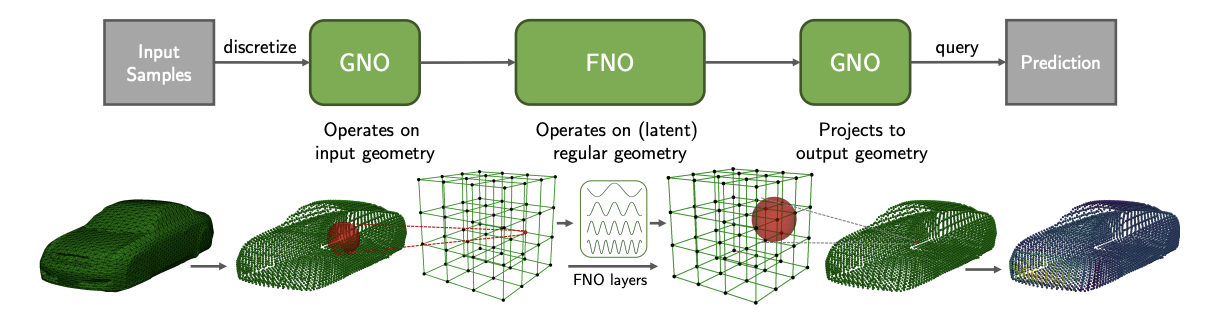}
    \caption{Architecture Diagram of GINO from \cite{gino}.}
    \label{fig:gino}
\end{figure}

Our implementation follows the original GINO formulation proposed in \cite{gino} (Figure \ref{fig:gino}), adapted to our application domain. GINO follows an encode–process–decode design, where information defined on irregular mesh nodes is first lifted into a latent regular grid representation, processed using a Fourier Neural Operator (FNO), and finally projected back onto the original mesh topology. 
In our implementation, the core encode–FNO–decode structure and geometric aggregation mechanism remain unchanged; the modifications are limited to the input feature construction and task-specific output projection. In fact, following our task formulation, each node feature vector is augmented with stimulus-relative positional information and four additional geometric features (three Cartesian coordinates, and the absolute distance between the stimulus point and the node). The configuration utilised in the experiments includes: a latent channel dimension of 48, a $12^3$ latent grid resolution, 6 Fourier modes, 2 FNO layers and a 100mm GNO radius (max. neighbors=16). 

\subsubsection{GWNO}
Global Wavelet Neural Operator (GWNO), is a neural operator architecture derived from GINO. The only difference between the two is the Fourier Neural Operator (FNO) core operator, as in GWNO, it is represented by a Wavelet Neural Operator (WNO) backbone (in place of a 3D FNO). The respective global operator definitions are formulated as follows: 

\begin{equation}
\begin{aligned}
GINO: K(u) = F^{-1} (R \cdot F(u)) &&
GWNO: K(u) = \Psi^{-1}(R_w \cdot \Psi(u))
\end{aligned}
\label{eq:ginogwno}
\end{equation}

where $\Psi$ and $\Psi^{-1}$ denote the discrete wavelet transform (DWT) and inverse DWT, respectively.

GWNO's architecture follows an encode–process–decode design and encompasses four stages: input lifting, GNO encoding, global operator propagation, and GNO decoding followed by projection.
Given an input function sampled on an irregular point cloud, the GNO encoder first maps the input coordinates and features into a latent representation on a regular grid, to then process the latent representation by the wavelet operator core. Finally, the decoded grid features are mapped back to the original query points through a GNO decoder and a projection MLP.

\section{Dataset and benchmark reproducibility}
\subsection{Dataset composition}
The 732 released simulations come from 20 left-ventricular anatomies extracted from the \texttt{zen4590294} four-chamber cohort~\cite{strocchi4chamber}, each simulated under a healthy substrate and three severities of random-walker scar, and paced from multiple Halton-sampled stimulus sites. Sample counts are unequal across substrates because per-run quality control removes simulations with degenerate or non-finite $V_m$; the frozen split records the exact surviving set.
\begin{table*}[htbp]
\centering
\scriptsize
\setlength{\tabcolsep}{3pt}
\renewcommand{\arraystretch}{1.08}

\caption{
Physiological calibration and beat-level quality assessment of the GraphEP-LV release ($n=732$). A \emph{quality beat} (QB) starts from diastole
($\leq1\%$ of nodes above $-70\,\mathrm{mV}$ at onset) and captures at least $95\%$ of eligible nodes within the pacing cycle.
Panel B reports QB trains for samples with a Purkinje tree and a diastolic initial state at $t=0$, thereby isolating the effect of lesion severity.
Panel C reports physiology over all samples contributing at least one QB; values are median [5th, 95th percentile]. CV and AT denote conduction velocity and activation time, respectively.}
\label{tab:graphep-lv-qc}

\begin{tabular*}{\linewidth}{@{\extracolsep{\fill}}llccc@{}}
\toprule
\multicolumn{5}{c}{\textbf{A. Conduction-velocity calibration}} \\
\midrule

\textbf{Tissue} & \textbf{Direction} & \textbf{Target CV} & \textbf{Achieved CV} & \textbf{Residual} \\

& & \textbf{(m\,s$^{-1}$)} & \textbf{(m\,s$^{-1}$)} & \textbf{(\%)} \\
\midrule

Myocardium (endo)
& Longitudinal & 0.67 & 0.658 & $-1.7$ \\

Myocardium (endo)
& Transverse & 0.30 & 0.310 & $+3.33$ \\

Myocardium (endo)
& Sheet-normal & 0.17 & 0.162 & $-4.70$ \\

Purkinje
& Longitudinal & 6.70 & 6.440 & $-3.88$ \\

Purkinje
& Transverse & 3.00 & 2.72 & $-9.33$ \\
\end{tabular*}
\par\vspace{0.5ex}
\begin{tabular*}{\linewidth}{@{\extracolsep{\fill}}lrccccc@{}}
\toprule
\multicolumn{7}{c}{
\textbf{B. Quality-beat train by severity
(Purkinje present, diastolic at $t=0$)}
} \\
\midrule

\textbf{Severity}
& \textbf{$n$} & \textbf{Mean QBs} & \textbf{0 QB} & \textbf{1 QB} & \textbf{2 QBs} & \textbf{3 QBs} \\

& & & \textbf{(\%)} & \textbf{(\%)} & \textbf{(\%)} & \textbf{(\%)} \\
\midrule

Healthy
& 48 & 2.94 & 0 & 0 & 6 & 94 \\

Minor
& 68 & 2.56 & 0 & 13 & 18 & 69 \\

Medium
& 59 & 1.22 & 2 & 85 & 3 & 10 \\

Large
& 54 & 0.89 & 11 & 89 & 0 & 0 \\
\end{tabular*}
\par\vspace{0.5ex}
\begin{tabular*}{\linewidth}{@{\extracolsep{\fill}}lrccccc@{}}
\toprule
\multicolumn{7}{c}{
\textbf{C. Physiology on quality beats}
} \\
\midrule

\textbf{Group}
& \textbf{$n$} & \textbf{Resting $V_m$} & \textbf{Amp.} & \textbf{APD$_{90}$} & \textbf{Activated} & \textbf{p95 AT} \\
& & \textbf{(mV)} & \textbf{(mV)} & \textbf{(ms)} & \textbf{(\%)} & \textbf{(ms)} \\
\midrule

All
& 269 & $-88.5\, [-88.8,-88.1]$ & $110.9\, [109.2,114.4]$ & $312\, [289,340]$ & $100.0\, [96.4,100.0]$ & $281\, [192,532]$ \\

Healthy
& 48 & $-88.1\, [-88.4,-88.0]$ & $113.7\, [111.7,115.0]$ & $300\, [293,316]$ & $100.0\, [100.0,100.0]$ & $237\, [181,282]$ \\

Minor
& 81 & $-88.3\, [-88.8,-88.1]$ & $111.7\, [111.1,115.1]$ & $312\, [296,337]$ & $100.0\, [99.9,100.0]$ & $234\, [184,412]$ \\

Medium
& 82 & $-88.8\, [-88.8,-88.3]$ & $109.7\, [109.4,111.2]$ & $318\, [285,347]$ & $99.9\, [95.7,100.0]$ & $327\, [228,557]$ \\

Large
& 58 & $-88.8\, [-88.8,-88.8]$ & $109.3\, [109.0,109.7]$ & $315\, [291,330]$ & $99.3\, [95.8,99.9]$ & $422\, [303,517]$ \\
\end{tabular*}
\par\vspace{0.5ex}
\begin{tabular*}{\linewidth}{@{\extracolsep{\fill}}lc@{}}
\toprule
\multicolumn{2}{c}{\textbf{D. Corpus composition}} \\
\midrule

\textbf{Purkinje tree present}
& $299/732$ (40.8\%) \\

\textbf{Diastolic at $t=0$}
& $441/732$ (60.2\%) \\
\bottomrule
\end{tabular*}

\end{table*}

\begin{table}[htbp]
\centering
\floatconts{tab:supp-corpus}%
{\caption{GraphEP corpus at a glance.  The benchmark predicts and evaluates the complete three-beat trajectory (361 frames; released graphs are \texttt{kmis}-coarsened from the full-resolution finite-element mesh.}}%
{\small
\begin{tabular}{@{}ll@{}}\toprule
Property & Value \\\midrule
Source cohort & \texttt{zen4590294} CMR meshes~\cite{strocchi4chamber} \\
LV anatomies & 20 \\
Substrate conditions & healthy; minor / medium / large random-walker scar \\
Simulations (H / Mi / Me / L) & 732 \; (100 / 220 / 212 / 200) \\
Pacing & 3 beats, BCL $600\,\mathrm{ms}$, $0$--$1800\,\mathrm{ms}$ \\
Stored / predicted frames & 361 at $5\,\mathrm{ms}$ over $0$--$1800\,\mathrm{ms}$ \\
Node / edge feature channels & 17 / 5 \\
Numerical precision & $V_m$, stimulus FP16; geometry, features FP32 \\
Train / val / test & 513 / 146 / 73 (stratified, anatomy-shared) \\
Release size & $39.85\,\mathrm{GB}$ across 11 files (9 data + metadata + manifest) \\
Estimated CPU hours & $2.6\times 10^4$ \\
\bottomrule\end{tabular}}
\end{table}

\subsection{Evaluation metrics and loss function}
\label{sec:supp-metrics}

\paragraph{Notation.}
One test sample is one simulation. For that sample let $V_{i,t}$ be the true
transmembrane voltage in mV at node $i=1,\dots,N$ and frame $t=1,\dots,T$, and
let $\hat V_{i,t}$ be the prediction. Frames are equally spaced by
$\Delta t = 5$ ms. Let $\mathcal{A}$ be the set of active-myocardium nodes (all nodes that
are not core scar) and $|\mathcal{A}|$ its size. Every metric below is computed on one
sample; the table reports the mean over the 73 test samples plus or minus the
standard deviation across those samples.

\paragraph{Voltage error (columns MSE, MAE, A-MAE).}
\begin{equation}
\mathrm{MSE}=\frac{1}{NT}\sum_{i=1}^{N}\sum_{t=1}^{T}\left(\hat V_{i,t}-V_{i,t}\right)^{2}
\end{equation}
\begin{equation}
\mathrm{MAE}=\frac{1}{NT}\sum_{i=1}^{N}\sum_{t=1}^{T}\left|\hat V_{i,t}-V_{i,t}\right|
\end{equation}
\begin{equation}
\text{A-MAE}=\frac{1}{|\mathcal{A}|\,T}\sum_{i\in \mathcal{A}}\sum_{t=1}^{T}\left|\hat V_{i,t}-V_{i,t}\right|
\end{equation}
A-MAE is the same average restricted to active myocardium, so that the passive
scar volume cannot dilute the error.

\paragraph{Depolarization amplitude (columns AR, DA-MAE).}
The amplitude of a node is its peak voltage above its own first frame,
\begin{equation}
A_i=\max_{t} V_{i,t}-V_{i,1},
\qquad
\hat A_i=\max_{t}\hat V_{i,t}-\hat V_{i,1} .
\end{equation}
\begin{equation}
\mathrm{AR}=\frac{\sum_{i=1}^{N}\hat A_i}{\sum_{i=1}^{N}A_i}
\end{equation}
\begin{equation}
\text{DA-MAE}=\frac{1}{N}\sum_{i=1}^{N}\left|\hat A_i-A_i\right|
\end{equation}
AR is a single ratio of mean amplitudes: $1$ means the action-potential swing is
reproduced, below $1$ means a damped prediction, above $1$ an over-shooting one.

\paragraph{Timing (columns LAT, APD90, RT).}
The activation time $\text{act}_i$ is the first time the trace rises through
$-40$ mV, obtained by linear interpolation between the two neighbouring frames.
The repolarization time $\text{repol}_i$ is the first time after the peak at which the
trace falls back to $V_{i,1}+0.1\,(\max_t V_{i,t}-V_{i,1})$, that is $90\%$
repolarized; $\hat{\text{act}}_i$ and $\hat{\text{repol}}_i$ are the same quantities of the
prediction. Let $S$ be the set of nodes that activate in the ground truth and
$R$ the set that also repolarizes in the ground truth.
\begin{equation}
\mathrm{LAT}=\frac{1}{|S|}\sum_{i\in S}\left|\hat{\text{act}}_i-\text{act}_i\right|
\end{equation}
\begin{equation}
\mathrm{APD90}=\frac{1}{|R|}\sum_{i\in R}\left|(\hat{\text{repol}}_i-\hat{ \text{act}}_i)-(\text{repol}_i-\text{act}_i)\right|
\end{equation}
\begin{equation}
\mathrm{RT}=\frac{1}{|R|}\sum_{i\in R}\left|\hat{\text{repol}}_i-\text{repol}_i\right|
\end{equation}
Timing is scored only where the ground truth provides the event. When a
prediction never crosses the threshold, its time is set to the last frame of the
window, so a missed activation or repolarization is counted as a large error
rather than silently dropped.

\paragraph{Training objective.}
Training minimizes a pointwise voltage term plus an amplitude guard. Let $\omega_i$
be the lumped node volume of node $i$, and let $\nu_i$ and $\hat \nu_i$ be the
standard deviation over time of the true and predicted trace of node $i$. For
one graph,
\begin{equation}
\mathcal{L}_{\text{MSE}}
=\frac{\sum_{i=1}^{N}\sum_{t=1}^{T} \omega_i\left(\hat V_{i,t}-V_{i,t}\right)^{2}}
       {T\sum_{i=1}^{N} \omega_i},
\qquad
\mathcal{L}_{\text{amp}}
=\frac{\sum_{i=1}^{N} \omega_i\left(\hat \nu_i-\nu_i\right)^{2}}{\sum_{i=1}^{N} \omega_i},
\end{equation}
\begin{equation}
\mathcal{L}=\mathcal{L}_{\text{MSE}}+0.1\,\mathcal{L}_{\text{amp}} .
\label{eq:supp-loss}
\end{equation}
The loss of a batch is the unweighted mean of $\mathcal{L}$ over its graphs, so
that large and small anatomies contribute equally. The amplitude guard is needed
because a plain voltage MSE is minimized by a damped, nearly flat trace that
averages the error across frames; penalizing the mismatch of the per-node
temporal standard deviation keeps the action-potential swing.

Two remarks. First, the metrics above are unweighted node averages, while the
loss is volume weighted; the two therefore agree in ranking but not in absolute
value. Second, the lag-based model additionally supervises its explicit
activation-time head against the baked ground-truth activation times with a
Huber term of transition width $5$ ms and weight $1$, added to
Equation~\eqref{eq:supp-loss}.

\subsection{Physical-volume-weighted sensitivity analysis}
\label{sec:supp-volume-metrics}
Uniform node averages can depend on spatial sampling density within a simulation. We therefore report a paired sensitivity analysis using the physical tissue volume represented by each pooled graph node. On the original tetrahedral mesh, the lumped $P_1$ volume at a vertex is one quarter of the sum of incident tetrahedron volumes. For the released KMIS graph, vertex volumes are summed over the exact saved pooling clusters to give weights $w_i$ for each retained node. Nodes with no represented tetrahedral tissue have zero weight. For each simulation, the volume-weighted voltage MAE is 
\begin{equation}
\mathrm{MAE}_{\mathrm{vol}}=
\frac{\sum_i w_i\ \sum_t|\hat V_{i,t}-V_{i,t}|}
     {T\sum_i w_i}.
\end{equation}
Where $T$ represent the amount of time frames. 

Volume-weighted LAT, APD90, and RT are computed under the same conditions of the table in the main section of the manuscript. We first calculate each metric per simulation and then report its mean and sample standard deviation across simulations. 

The analysis covers all seven Table~1 models with paired scores. The volume-weighted MAE ordering is unchanged: GAP$^2$ODE retains the lowest voltage MAE, and HGT-Lag retains the lowest APD90 and RT errors. Weighting reverses three close secondary comparisons: GWNO versus HGT on APD90, GHR versus GINO on RT, and GWNO versus GAP$^2$ODE on RT. 

\begin{table*}[t]
\centering
\small
\caption{Paired node- and physical-volume-weighted test metrics across the seven available models (73 simulations per model). Values are mean $\pm$ sample standard deviation across simulations. Volume weights are summed over exact KMIS clusters; voltage metrics score the released pooled graph field.}
\label{tab:volume-weighted}
\resizebox{\linewidth}{!}{
\begin{tabular}{lrrrrr}
\hline
Model & Node MAE & Volume MAE & Volume LAT & Volume APD90 & Volume RT \\
 & (mV) & (mV) & (ms) & (ms) & (ms) \\
\hline
GATv2 & $35.33 \pm 4.29$ & $35.71 \pm 4.33$ & $1235.10 \pm 241.93$ & $600.42 \pm 110.48$ & $653.98 \pm 234.74$ \\
GHR & $20.31 \pm 5.65$ & $19.90 \pm 6.09$ & $272.56 \pm 183.60$ & $471.11 \pm 123.79$ & $439.10 \pm 184.62$ \\
GINO & $20.78 \pm 8.27$ & $20.58 \pm 8.42$ & $188.59 \pm 180.69$ & $476.35 \pm 153.31$ & $440.41 \pm 190.27$ \\
GWNO & $20.54 \pm 8.26$ & $20.32 \pm 8.41$ & $178.00 \pm 173.41$ & $425.40 \pm 118.24$ & $405.80 \pm 167.16$ \\
HGT & $15.55 \pm 5.09$ & $15.42 \pm 5.24$ & $175.35 \pm 180.07$ & $429.18 \pm 107.86$ & $399.70 \pm 154.66$ \\
HGT-Lag & $16.86 \pm 7.61$ & $16.85 \pm 7.85$ & $241.67 \pm 265.74$ & $368.64 \pm 151.98$ & $279.12 \pm 119.69$ \\
GAP2ODE & $10.21 \pm 4.62$ & $10.16 \pm 4.80$ & $98.10 \pm 122.56$ & $436.63 \pm 128.77$ & $404.82 \pm 163.31$ \\
\hline
\end{tabular}
}

\end{table*}

\subsection{Examples of predicted samples}
\label{sec:supp-prediction-examples}
\begin{figure}
    \centering
    \includegraphics[width=1\linewidth]{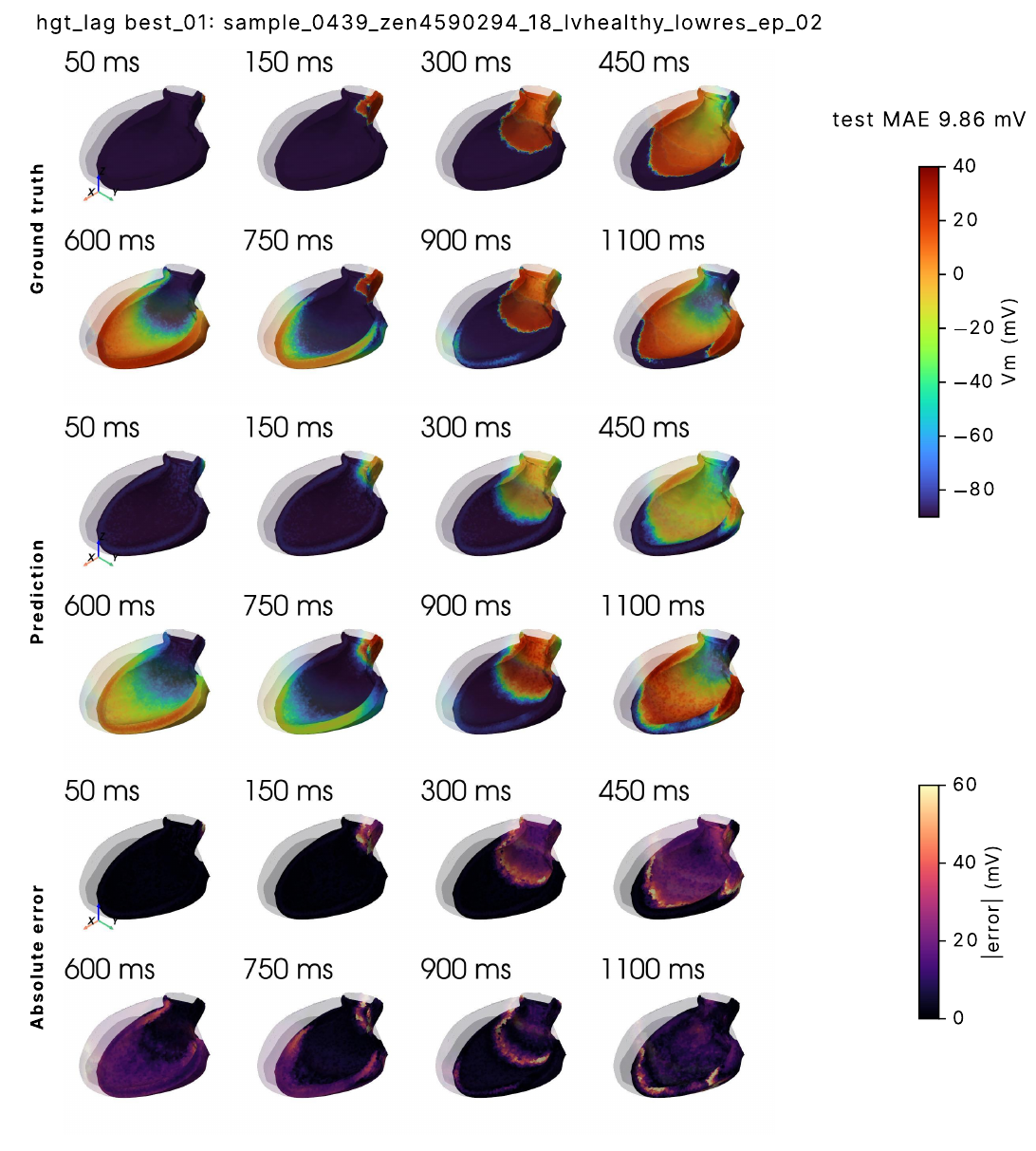}
    \caption{These are half cut spatial renders of different temporal snapshots of an healthy prediction. The figure reports in the first row the ground truth target; on the second the prediction of the model; in the third row the absolute error. Colormap is related to the signal and provided by the colorbar on the right. }
    \label{fig:healthy-prediction}
\end{figure}

\begin{figure}
    \centering
    \includegraphics[width=1\linewidth]{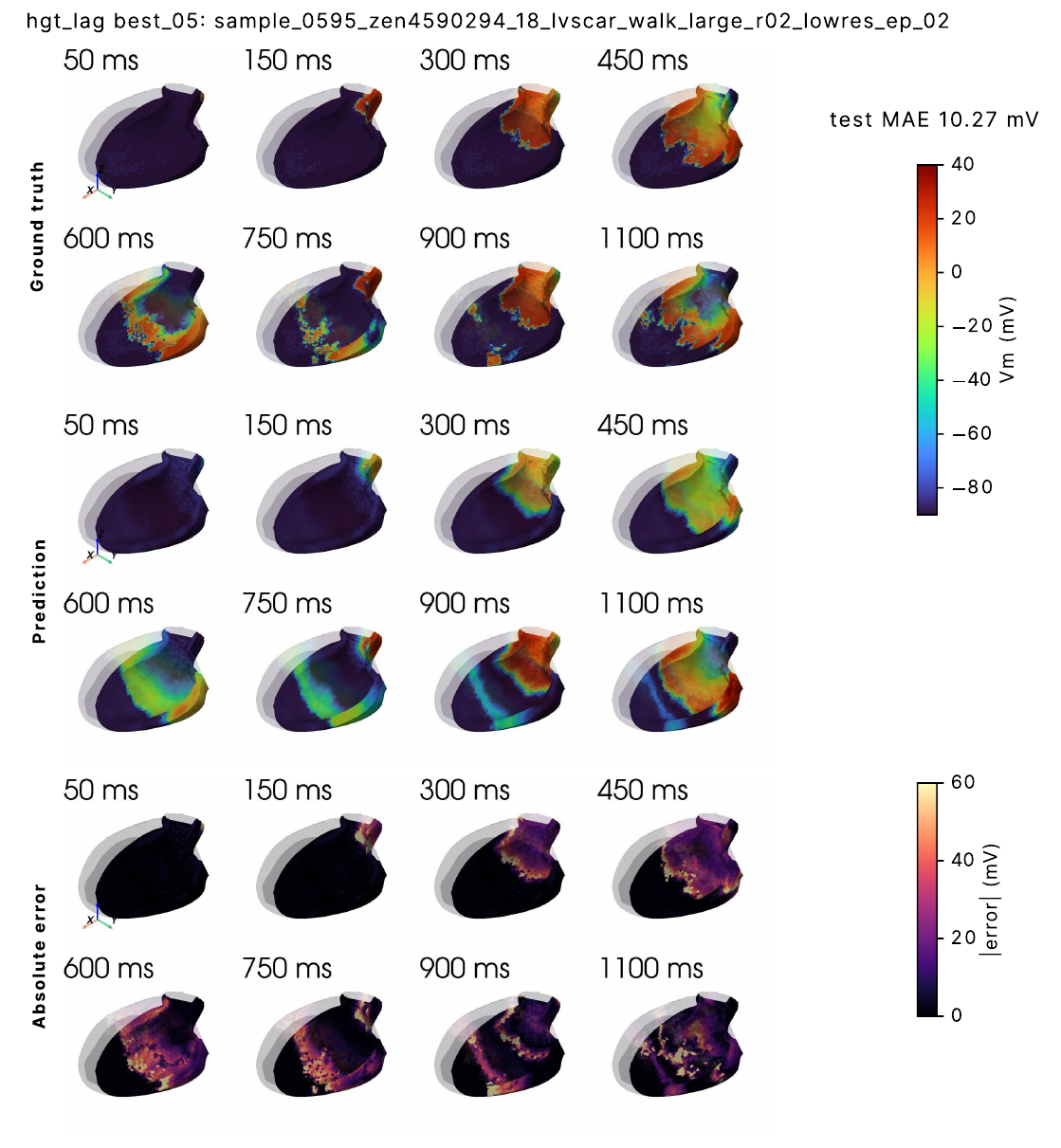}
    \caption{These are half cut spatial renders of different temporal snapshots of a scarred prediction. The figure reports in the first row the ground truth target; on the second the prediction of the model; in the third row the absolute error. Colormap is related to the signal and provided by the colorbar on the right. The scar produces delay and there are signals overlapping.}
    \label{fig:scar-prediction}
\end{figure}

In this subsection we report a scar and healthy example with high performance from the propagation-aware in Figures [\ref{fig:healthy-prediction},\ref{fig:scar-prediction}].


\section{Other examples}
\subsection{Biventricular and whole-heart runs}
\label{sec:supp-multichamber-reproduction}
\begin{figure}
    \centering
    \includegraphics[width=\linewidth]{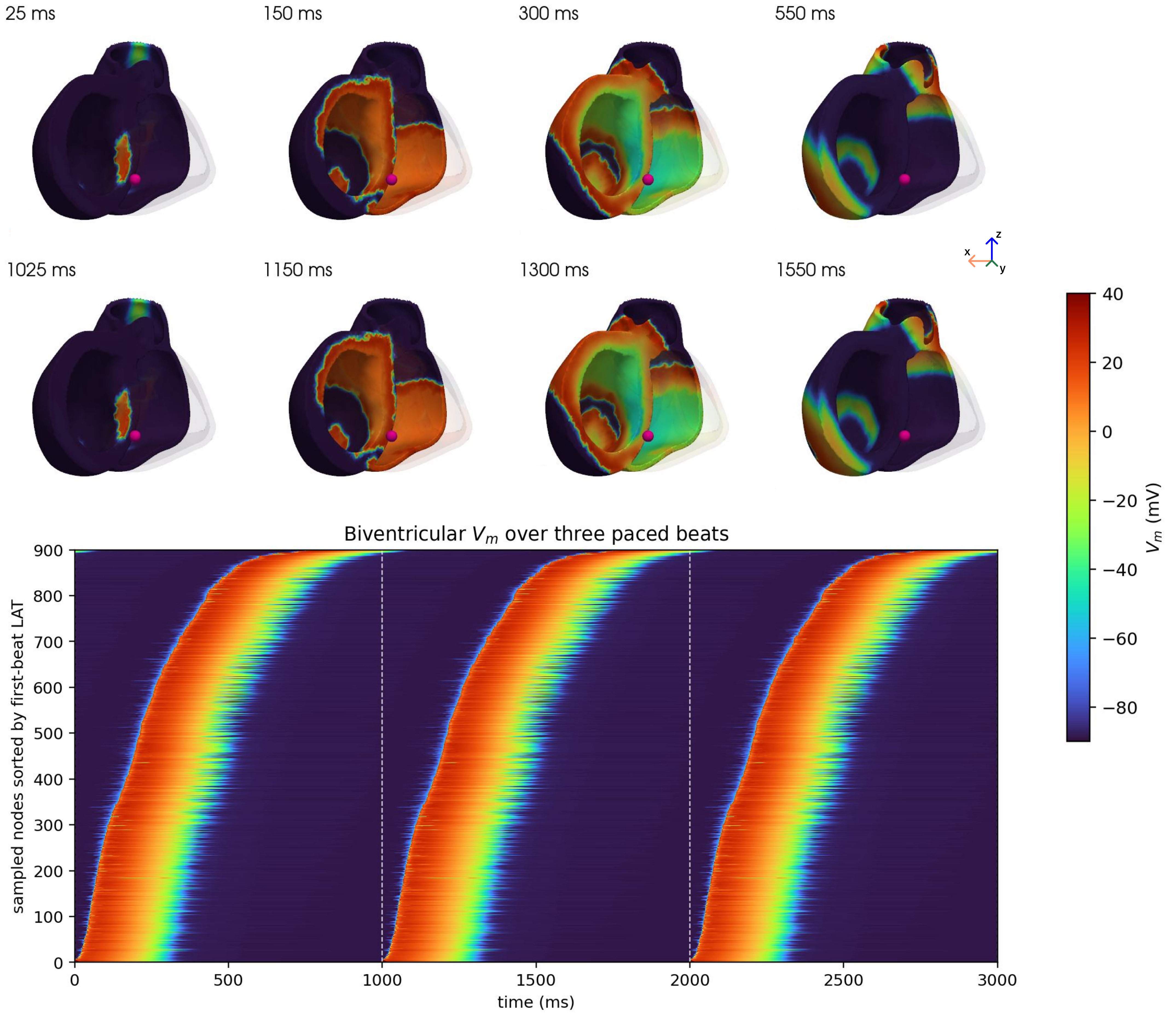}
    \caption{epresentative healthy biventricular simulation shown at six stages of the paced trajectory. Colour denotes transmembrane voltage. This example demonstrates extension of the generation pipeline to coupled LV and RV tissue with chamber-specific fast-conduction regions; it is not part of the quantitative LV benchmark.}
    \label{fig:supp-biv-example}
\end{figure}

\begin{figure}
    \centering
    \includegraphics[width=\linewidth]{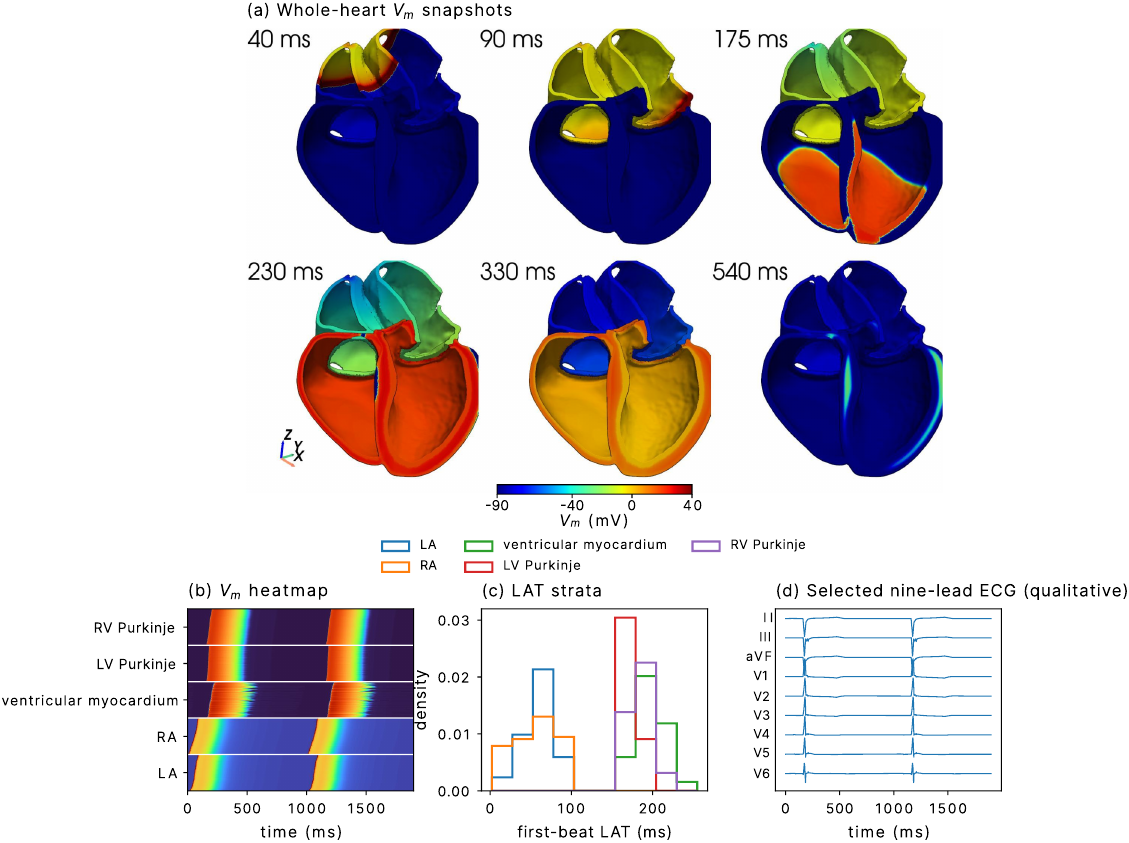}
    \caption{Representative healthy whole-heart simulation shown at six stages of the activation trajectory. Colour denotes transmembrane voltage. The example couples atrial and ventricular electrophysiology and is included as a pipeline demonstration rather than part of the quantitative LV benchmark.}
    \label{fig:supp-wholeheart-example}
\end{figure}

Extended Tables~\ref{tab:supp-multichamber-pipeline} and
\ref{tab:supp-multichamber-tissue} classify the parameters of the healthy
biventricular and whole-heart examples by computational-pipeline stage.
These are the resolved plans used to produce Figures~\ref{fig:supp-biv-example} and \ref{fig:supp-wholeheart-example}, rather than generic defaults. Coordinates and radii are in $\mu$m, times in ms, conduction velocities in m/s, and conductivities in S/m. Both runs use anatomy 01 from the \texttt{zen4590294} four-chamber collection.

\begin{table}[htbp]
\floatconts{tab:supp-multichamber-pipeline}%
{\caption{Taxonomy of the parameters needed to reproduce the healthy multichamber runs.  $f$, $s$ and $n$ denote the fibre, sheet and sheet-normal directions. }}%
{{\scriptsize\centering
\setlength{\tabcolsep}{3pt}
\renewcommand{\arraystretch}{1.12}
\resizebox{\linewidth}{!}{%
\begin{tabular}{@{}
>{\raggedright\arraybackslash}p{0.12\linewidth}
>{\raggedright\arraybackslash}p{0.20\linewidth}
>{\raggedright\arraybackslash}p{0.31\linewidth}
>{\raggedright\arraybackslash}p{0.31\linewidth}
@{}}\toprule
Pipeline stage & Parameter class & Biventricular healthy & Whole-heart healthy \\\midrule
Geometry & Chamber extraction & Source tags 1--2 (LV, RV) & Source tags 1--4 (LV, RV, LA, RA) \\
 & Resolution & target 2000; allowed 1500--3000 & target X; allowed X--X \\
 & Coordinates/fibres & VTK conversion; interpolated UVC and source fibre field & Same \\
Tissue tagging & Ventricular wall & UVC $\rho$: endo $[0,.25)$, mid $[.25,.75)$, epi $[.75,1]$ & Same; source tags additionally separate LA and RA \\
 & Purkinje construction & LV/RV adaptive trees; $N_{it}=22$ (maximum 26); apex roots; seeds 1213/2213 & Initial LV/RV trees as at left; LV root \texttt{random\_endo}, RV root apex; subsequently converted to a line-network representation \\
 & Purkinje coverage & target 0.30, radius 1200, one retry, distance-threshold factor 0.5 & Same as biventricular model \\
Cell initialization & Ionic steady state & Tomek with default; 50 single-cell cycles at BCL 1000 & Same for the ventricles; Courtemanche LA/RA; 50 cycles at BCL 1000 \\
CV calibration & Target $v_f/v_s/v_n$ & myocardium 0.67/0.30/0.17; Purkinje 6.7/3.0/1.7 & Same; atria 0.80/0.40/0.20 \\
Full-mesh prepacing & Protocol & 5 stimulation cycles, BCL 1000; simultaneous LV/RV stimuli & 5 cycles, BCL 1000; SAN at 0, LV/RV roots at 160 \\
Production EP & Stimulus support & LV/RV spheres, radius 5000, centres $(71783,63136,26527)$ and $(108845,15599,13324)$ & SAN sphere, radius 5000, centre $(17537,56609,102672)$; LV/RV root vertices 60159/86326;\\
 & Timing & 3 pulses, start 0, duration 2, BCL 1000; strength 250 & 3 pulses per site, duration 2, BCL 1000; SAN/root offsets 0/160;) \\
Numerics & Monodomain solver & $dx=250$, $dt=1$, tolerance $10^{-5}$, operator splitting and mass lumping; $\texttt{spacedt}=0.5$, $\texttt{timedt}=1$ & Same \\
\bottomrule\end{tabular}%
}}}
\end{table}

The resulting effective material parameters are reported separately in
Table~\ref{tab:supp-multichamber-tissue}.  Values are ordered as
$(f,s,n)$.  The biventricular values and the unscaled whole-heart values were
obtained by the calibration described above.

\begin{table}[htbp]
\centering
\floatconts{tab:supp-multichamber-tissue}%
{\caption{Ionic models and effective orthotropic conductivities used in the
resolved simulation plans. The biventricular values were obtained by
CV-targeted calibration from literature-informed seeds; the whole-heart
ventricular values include the documented post-calibration factor of nine. The
intracellular and extracellular triples are denoted by $g_i$ and $g_e$,
respectively.  ``Both'' means that the same values were used in the
biventricular and whole-heart examples.}}%
{{\scriptsize\centering
\setlength{\tabcolsep}{4pt}
\begin{tabular}{@{}lllll@{}}\toprule
Run & Tissue & Ionic model & $g_i=(g_{if},g_{is},g_{in})$ & $g_e=(g_{ef},g_{es},g_{en})$ \\\midrule
Biventricular & endocardium & Tomek ENDO & (0.23048, 0.06387, 0.02906) & (0.62, 0.24, 0.24) \\
 & mid-myocardium & Tomek MCELL & (0.23713, 0.06531, 0.02956) & (0.62, 0.24, 0.24) \\
 & epicardium/default myocardium & Tomek EPI & (0.23324, 0.06494, 0.02978) & (0.62, 0.24, 0.24) \\
Whole heart & endocardium & Tomek ENDO & (2.07434, 0.57487, 0.26154) & (5.58, 2.16, 2.16) \\
 & mid-myocardium & Tomek MCELL & (2.13419, 0.58776, 0.26605) & (5.58, 2.16, 2.16) \\
 & epicardium/default myocardium & Tomek EPI & (2.09919, 0.58442, 0.26804) & (5.58, 2.16, 2.16) \\
 & left/right atrium & Courtemanche & (3.50587, 1.06203, 0.37812) & (5.58, 2.16, 2.16) \\
Both & LV/RV Purkinje & Tomek ENDO & (18.36735, 3.77421, 1.25709) & (62.0, 24.0, 24.0) \\
\bottomrule\end{tabular}}}
\end{table}

\subsection{Extended results of the benchmark}
\begin{figure}[htbp]
      \centering
      \includegraphics[width=\linewidth]{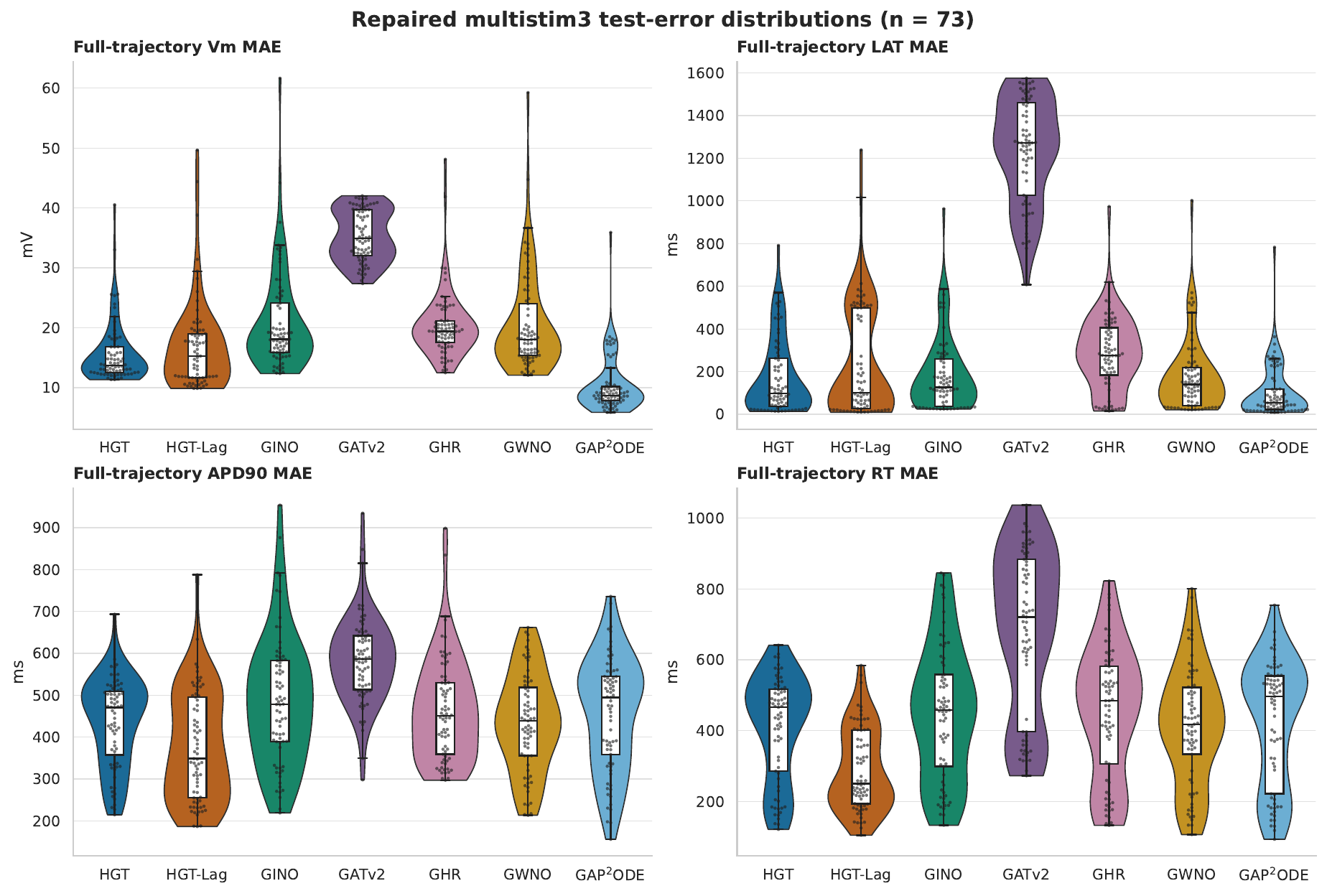}
      \caption{Full-trajectory test-error distributions across the 73 held-out simulations. Violins show the empirical sample distribution, internal boxes show the median and interquartile range, and points represent individual simulations. All panels use the 361-frame targets.}
      \label{fig:model-capability-distributions}
\end{figure}

In this subsection we report the statistical distribution over the samples metrics of the models on the test set. It is increasingly evident that HGT-lag outperforms the other models in detectng the waveform of the signal, even if overall makes more mistakes on some samples that it finds harder. In fact it tests on a subset of samples with the highest performance, as stated in Figure \ref{fig:model-capability-distributions}

\end{document}